\documentclass[preprintnumbers,floatfix,aps,prd,preprint]{revtex4-1}

\usepackage{amsmath}
\usepackage{amsfonts}
\usepackage{amssymb}
\usepackage{bm}
\usepackage{enumerate}
\usepackage{color}
\usepackage{graphicx}
\usepackage{epsfig}
\usepackage[figuresright]{rotating}
\usepackage{hhline,multirow}  
\usepackage{dcolumn}          
\usepackage[colorlinks=true,linkcolor=blue]{hyperref}
\usepackage{color,soul}
\newcommand{\eps}{\varepsilon}

\begin{document}
\preprint{JLAB-THY-18-2865}

\title{\mbox{Deep-inelastic and quasielastic electron scattering
		from $A=3$ nuclei}}

\author{A.~J.~Tropiano$^1$,
	J.~J.~Ethier$^2$,
	W.~Melnitchouk$^3$,
	N.~Sato$^{4,5}$}
\affiliation{
$^1$\mbox{The Ohio State University, Columbus, Ohio 43210, USA} \\
$^2$\mbox{Department of Physics and Astronomy,
	  Vrije Universiteit Amsterdam,
	  1081 HV Amsterdam},	\\
    \mbox{and Nikhef Theory Group, Science Park 105, 1098 XG Amsterdam,
	  The Netherlands} \\
$^3$\mbox{Jefferson Lab, Newport News, Virginia 23606, USA}	\\
$^4$\mbox{University of Connecticut, Storrs, Connecticut 06269, USA} \\
$^5$\mbox{Old Dominion University, Norfolk, Virginia 23529, USA}
}

\begin{abstract}
We perform a combined analysis of inclusive electron scattering data
from $A=3$ nuclei in the deep-inelastic and quasielastic scattering
regions, using Monte Carlo analysis methods and the nuclear weak
binding approximation to establish the range over which the data
can be described within the same theoretical framework.
Comparison with quasielastic $^3$He cross sections from SLAC and
Jefferson Lab suggests that most features of the $x \gtrsim 1$ data
can be reasonably well described in the impulse approximation with
finite-$Q^2$ nuclear smearing functions for momentum transfers
$Q^2 \gtrsim 1$~GeV$^2$.
For the DIS region, we analyze the recent $^3$He to deuterium
cross section ratio from the Jefferson Lab E03-103 experiment to
explore the possible isospin dependence of the nuclear effects.
We discuss the implications of this for the MARATHON experiment
at Jefferson Lab, and outline how a Bayesian analysis of $^3$He,
$^3$H and deuterium data can robustly determine the free neutron
structure function.
\end{abstract}

\date{\today}
\maketitle

\section{Introduction}
\label{sec:intro}

With the completion of the 12~GeV energy upgrade of Jefferson Lab,
a new chapter in the exploration of the quark structure of the nucleon
has begun.  One of the main drivers of the new facility is the
determination of the spatial, momentum, and spin distributions of the
nucleon's valence quarks.  Of particular interest are configurations
in which a single quark carries a large fraction, $x$, of the momentum
of the nucleon, which can reveal details of the underlying quark-gluon
dynamics~\cite{Holt:2010vj}.

It is surprising that, almost four decades after the first
experimental deep-inelastic scattering (DIS) programs were initiated,
such fundamental quantities as the momentum fraction carried by $d$
quarks in the proton are still poorly known at large
$x$~\cite{JLab11, JMO13, Gao:2017yyd}.
While this is partly due to the steeply falling inclusive DIS rates
as $x \to 1$, the additional complication has been the absence of
free neutron targets, which has significantly limited the extraction
of $u$ and $d$ flavor information from hydrogen and deuterium data
due to nuclear effects in the latter~\cite{MT96}.
Indeed, uncertainties from the short-range part of the nucleon--nucleon
interaction give rise to differences in the extracted $d/u$ parton
distribution function (PDF) ratio that are typically of the same order
as the variation between predictions from different dynamical
models~\cite{CJ11, CJ12, Rubin12}.

Recent progress on the experimental front has come with the
measurement of the nearly-free neutron structure function in the
``BONuS'' experiment at Jefferson Lab~\cite{Baillie12, Tkachenko14},
using spectator tagging in semi-inclusive DIS from the deuteron,
which has improved the precision of the $d/u$ ratio in the
intermediate- to high-$x$ region.
More dramatically, data on charged lepton and $W$-boson asymmetries
in $p\bar p$ collisions from the CDF and D0 Collaborations at
Fermilab~\cite{CDF05, CDF09, D008l, D008e} have provided more
stringent constraints on the $d/u$ behavior up to $x \sim 0.7$.
In particular, the recent CJ15 global QCD analysis~\cite{CJ15}
suggested that the nucleon off-shell effects in the deuteron are
relatively small, at least in the isoscalar channel.

However, while the new data have led to a reduction in the extracted
PDF uncertainties at large $x$, there is still considerable
uncertainty in the extrapolation from the highest $x$ values
at which there data to the $x=1$ limit.  For instance,
depending on the functional form chosen for the $u$ and $d$ PDFs,
one can get rather different extrapolated $d/u$ ratios in the
$x=1$ limit~\cite{CJ15, AKP17}.
The experimental program at Jefferson Lab at 12~GeV aims to bridge
this gap by using several novel techniques to isolate the $d/u$
ratio up to $x \sim 0.85$ in the DIS region.
The spectator tagging method will be used again to extend the BONuS
experiment to 12~GeV~\cite{BONUS12}, isolating nearly free neutrons
in the deuteron by detecting a low-momentum, backward-angle proton
in DIS off deuterium.
Further ahead, the SoLID Collaboration aims to measure parity-violating
DIS from the proton, with the $\gamma$-$Z$ interference structure
function providing a different combination of the $u$ and $d$ PDFs
compared with electromagnetic scattering~\cite{SOLID, Hobbs08}.

In this paper we consider the alternative method proposed to extract
the $d/u$ ratio, using the measurement of DIS cross sections from
$^3$He and $^3$H nuclei with the MARATHON experiment at Jefferson
Lab~\cite{MARATHON}, which completed data taking in 2018.
It was shown in Refs.~\cite{Afnan00, Afnan03, Pace01, Sargsian02} that, 
under reasonable assumptions about the isospin dependence of nucleon
off-shell effects, the ratio of $^3$He to $^3$H structure functions
could directly constrain the neutron to proton ratio, $F_2^n/F_2^p$,
with nuclear effects largely canceling between the mirror nuclei.
From knowledge of the free neutron to proton ratio, one can then
directly extract $d/u$ in the valence quark dominated region,
$x \gtrsim 0.4$.

Since the earlier calculations, progress on the theoretical front
has been made in computing structure functions of light nuclei
within the framework of the weak binding approximation
(WBA)~\cite{KP06, KM08, Ethier13}, including finite-energy
corrections and nucleon off-shell contributions.
In the case of the DIS from the deuteron, the latter have been
estimated within nuclear models~\cite{CJ11, CJ12} and fitted in
phenomenological analyses~\cite{KP06, CJ15, AKP17} for a given
set of deuteron wave functions.
Information on the off-shell effects in $A=3$ nuclei, on the other
hand, has been more difficult to obtain, partly because of the dearth
of data on unpolarized $^3$He structure functions (and the complete
absence for $^3$H).  This had left open the possibility of
potentially large isovector off-shell effects~\cite{Sargsian02},
which would contribute to $^3$He/$^3$H structure functions,
but not be seen in DIS from deuterium.

In the present work we revisit the question of the isospin dependence
of off-shell effects in the light of more recent data from the
Jefferson Lab E03-103 experiment~\cite{Seely09}, which measured
ratios of structure functions of light nuclei to those of deuterium.
In particular, the experiment obtained the first high-precision
determination of the $^3$He to deuterium cross section ratio for
$x \sim 0.3-0.6$ in DIS kinematics.  These data have the potential
to constrain, when combined with the inclusive deuterium DIS data,
the individual off-shell corrections to the proton and neutron
structure functions, and clarify the impact on the extracted
$F_2^n/F_2^p$ ratio.
In Ref.~\cite{KP10}, for example, the data were used to benchmark
the $n/p$ ratio extracted from E03-103 with that obtained from
earlier inclusive proton and deuterium data sets, requiring a
``renormalization'' of the $^3$He to deuterium cross section
ratio by $+3\%$.
Here we re-examine the E03-103 $^3$He/deuterium data, in combination
with the isoscalar nucleon off-shell corrections obtained from the
recent CJ15 global QCD analysis \cite{CJ15}, and place upper limits
on the magnitude of the isospin dependence of the off-shell corrections.

To further constrain the models of the nuclear effects, we test
the efficacy of the $^3$He smearing functions computed within
the WBA framework to simultaneously describe other processes,
such as quasielastic (QE) electron scattering from $^3$He nuclei.
We compare with the available QE data from experiments at
SLAC~\cite{Day:1979bx, Meziani:1992xr} and
Jefferson Lab~\cite{Fomin:2010ei} in the region $x \gtrsim 1$ and
at four-momentum transfers $Q^2 \sim 1$ to a few GeV$^2$, where the
nuclear impulse approximation is expected to be valid.

We begin in Sec.~\ref{sec:theory} by reviewing the formalism for
inclusive lepton scattering from nuclei, and summarizing the results
for nuclear structure functions in terms of on-shell and off-shell
convolutions of nucleon structure functions and nucleon (light-cone)
momentum distribution functions in $A=3$ nuclei.
Here we also illustrate the specific features of the nucleon light
cone distributions (which are also referred to as nucleon
``smearing functions'') as a function of nuclear momentum fraction
and $Q^2$.
The versatility of the smearing functions in describing different
$^3$He observables is discussed in Sec.~\ref{sec:qe}, where we
compare the QE cross sections calculated in the WBA with data on
inclusive electron--$^3$He scattering in the QE region,
$x \sim 1$, from SLAC~\cite{Day:1979bx, Meziani:1992xr}
and Jefferson Lab~\cite{Fomin:2010ei}.
After establishing the kinematic regions in $x$ and $Q^2$ where
the data can be accommodated, we estimate the QE cross sections
for $^3$He and $^3$H at the kinematics of the E12-11-112
experiment at Jefferson Lab~\cite{E12-11-112}.

DIS from $^3$He and $^3$H nuclei is discussed in Sec.~\ref{sec:dis}.
Here we fit the recent 6~GeV Jefferson Lab data~\cite{Seely09} on
the $^3$He to deuterium cross section ratio to extract the isovector
component of the nucleon off-shell contributions.  We use several
different nuclear models and off-shell parametrizations to estimate
the theoretical uncertainty in the extracted off-shell corrections,
and determine the impact on the extraction of the $F_2^n/F_2^p$ ratio.
Finally, in Sec.~\ref{sec:conc} we summarize our findings and
anticipate future developments in experiment and theory which may
reveal further insight into both the quark structure of the nucleon
and the dynamics of $^3$He and $^3$H nuclei.

\section{Formalism}
\label{sec:theory}

In this section we summarize the basic formulas for inclusive
electron scattering from nuclei.  We present the results for the
nuclear structure functions in the framework of the WBA, in which the
structure functions of the nucleus are represented as convolutions
of nucleon momentum distributions in the nucleus and structure
functions of (off-shell) nucleons, up to ${\cal O}(\bm{p}^2/M^2)$
corrections, where $\bm{p}$ and $M$ are the 3-momentum and mass of
the initial state nucleon~\cite{KP06, KM08, KPW94, KMPW95, KM08d}.
(Higher order relativistic effects necessarily lead to a
breakdown of the factorization embodied in the convolution
representation~\cite{MST94, MST94plb}.)
After providing the complete set of formulas for structure functions
for scattering of both transverse and longitudinal photons,
we illustrate the smearing functions, for on-shell and off-shell
nucleon contributions, for $A=3$ nuclei.

\subsection{Inclusive nuclear cross section and structure functions}

We consider the inclusive scattering of an electron from a nucleus $A$
(later specializing to the case $A =$ $^3$He and $^3$H),
	$e A \to e X$,
where $X$ represents the unobserved hadronic state.
We denote the four-momenta of the incident and scattered electrons by
$k_\mu$ and $k'_\mu$, respectively, and the four-momentum of the target
by $P_\mu$.  In the target rest frame the inclusive cross section is
given by
\begin{eqnarray}
\frac{d^2\sigma}{d\Omega dE'}
&=& \frac{\alpha^2}{Q^4} \frac{E'}{E} \frac{1}{M_A}
    L_{\mu\nu}\, W^{\mu\nu},
\label{eq:sigLW}
\end{eqnarray}
where $\alpha$ is the fine structure constant, $E$ ($E'$) is the
energy of the incident (scattered) electron, and $M_A$ is the mass
of the nucleus.
The four-momentum of the exchanged virtual photon is
$q_\mu = k_\mu - k'_\mu$.  The invariant mass squared of the photon
can be approximated by neglecting the small electron mass,
	$Q^2 \equiv -q^2 \approx 4 E E' \sin^2(\theta/2)$,
where $\theta$ is the angle between the incident and scattered
electrons.
The leptonic tensor in Eq.~(\ref{eq:sigLW}) is given by
\begin{eqnarray}
L_{\mu\nu}
&=& 2 k_\mu k'_\nu + 2 k'_\mu k_\nu + q^2 g_{\mu\nu},
\end{eqnarray}
and the hadronic tensor is parametrized by the nuclear structure
functions $F_1^A$ and $F_2^A$,
\begin{eqnarray}
W^{\mu\nu}(P,q)
&=& \left( -g^{\mu\nu} + \frac{q^\mu q^\nu}{q^2} \right)
     F_1^A\
 +\ \left( P^\mu - \frac{P \cdot q}{q^2} q^\mu \right)
    \left( P^\nu - \frac{P \cdot q}{q^2} q^\nu \right)
    \frac{F_2^A}{P \cdot q}\, .
\label{eq:Wmunu_gen}
\end{eqnarray}
The structure functions are taken to be functions of $Q^2$
and the Bjorken scaling variable, $x = Q^2 / 2 M \nu$,
where $\nu = E - E'$ is the energy transfer.
One can then write the inclusive cross section in terms of the 
nuclear structure functions as
\begin{eqnarray}
\label{eq:sigma}
\sigma^A\ \equiv\
\frac{d^2\sigma}{d\Omega dE'}
&=& \sigma_{\rm Mott}
    \left( \frac{2}{M_A} \tan^2\frac{\theta}{2} F_1^A(x,Q^2)
	 + \frac{1}{\nu} F_2^A(x,Q^2)
    \right),
\end{eqnarray}
where $\sigma_{\rm Mott} = (4\alpha^2 E'^2/Q^4) \cos^2(\theta/2)$
is the Mott cross section for scattering from a point particle.
Note that for forward scattering, $\theta=0^\circ$, the cross section
is dominated by the $F_2^A$ structure function, while for backward
scattering, $\theta=180^\circ$, it is given only in terms of $F_1^A$.
For intermediate scattering angles, both the $F_1^A$ and $F_2^A$
structure functions contribute to the cross section.

Alternatively, one can also write the hadronic tensor and cross section
in terms of the transverse and longitudinal structure functions,
$F_T^A$ and $F_L^A$, corresponding to the contributions to the
scattering from exchanged photons with transverse or longitudinal
polarization, respectively,
\begin{subequations}
\begin{eqnarray}
F_T^A(x,Q^2) &=& 2 x F_1^A(x,Q^2),	\\
F_L^A(x,Q^2) &=& \gamma^2\, F_2^A(x,Q^2) - F_T^A(x,Q^2),
\end{eqnarray}
\end{subequations}
where the kinematical parameter
\begin{eqnarray}
\gamma^2 &\equiv& \frac{\bm{q}^2}{\nu^2}\,
              =\, 1 + \frac{4 M^2 x^2}{Q^2}
\label{eq:gamma}
\end{eqnarray}
accounts for finite-energy effects.
Note that sometimes in the literature one uses the nuclear scaling
variable, $x_A = (M/M_A)\, x$, which ranges between 0 and 1.
In the present analysis we will use the variable $x$ when
comparing structure functions of nuclei and nucleons.

\subsection{Structure functions in the weak binding approximation}
\label{ssec:wba}

Neglecting antinucleon degrees of freedom, in the WBA the nucleus is
approximated as a system of weakly bound nucleons with four-momentum
	\mbox{$p_\mu \equiv (M+\eps, \bm{p})$},
where the nucleon three-momentum $\bm{p}$ and off-shell energy
energy $\eps$ ($<0$) are both much smaller than the nucleon mass,
	$|\bm{p}|, |\eps|~\ll~M$~\cite{KPW94, KMPW95}.
Reducing the relativistic Lorentz-Dirac structures in the general
decomposition of the off-shell nucleon hadronic tensor~\cite{MST94,
KPW94}, one can relate the relativistic four-component nucleon
field to the corresponding two-component operator, up to order
${\cal O}(\bm{p}^2/M^2)$~\cite{KP06, KM08}.
The imaginary part of the nucleon propagator can then be written
in terms of a nuclear spectral function defined through the
correlator of the nonrelativistic fields.

A lengthy but straightforward derivation then allows one to show that
the nuclear structure functions can be written in factorized form,
\begin{subequations}
\label{eq:F1F2_gen}
\begin{eqnarray}
x F_1^A(x,Q^2)
&=& \sum_N
    \int\!\frac{d^4{p}}{(2\pi)^4}
    {\cal F}_0^N\left(\eps,{\bm p}\right)
    \left( 1 + \frac{\gamma p_z}{M} \right)	\nonumber\\
& & \hspace*{2cm} \times
    \left[ {\cal C}_{11}\,
	   \frac{x}{y} \widetilde{F}_1^N\left(\frac{x}{y},Q^2,p^2\right)
	 + {\cal C}_{12}\,
	   \widetilde{F}_2^N\left(\frac{x}{y},Q^2,p^2\right)
   \right],
\label{eq:F1_gen}				\\
F_2^A(x,Q^2)
&=& \sum_N
    \int\!\frac{d^4{p}}{(2\pi)^4}
    {\cal F}_0^N\left(\eps,{\bm p}\right)
    \left( 1 + \frac{\gamma p_z}{M} \right)
    {\cal C}_{22}\, \widetilde{F}_2^N\left(\frac{x}{y},Q^2,p^2\right),
\label{eq:F2_gen}				\\
F_L^A(x,Q^2)
&=& \sum_N
    \int\!\frac{d^4{p}}{(2\pi)^4}
    {\cal F}_0^N\left(\eps,{\bm p}\right)
    \left( 1 + \frac{\gamma p_z}{M} \right)	\nonumber\\
& & \hspace*{2cm} \times
    \left[ {\cal C}_{LL}\,
	   \widetilde{F}_L^N\left(\frac{x}{y},Q^2,p^2\right)
	 + {\cal C}_{L2}\,
	   \widetilde{F}_2^N\left(\frac{x}{y},Q^2,p^2\right)
   \right],
\label{eq:FL_gen}
\end{eqnarray}
\end{subequations}
where the sum is over nucleons $N = p, n$, the function ${\cal F}_0^N$
is the nonrelativistic nucleon spectral function in the nucleus,
and $\widetilde{F}_i^N$ $(i=1,2,L)$ are the off-shell nucleon structure
functions, which depend also on the nucleon virtuality, $p^2$.
The variable
\begin{eqnarray}
y &\equiv& \frac{M_A}{M} \frac{p \cdot q}{P \cdot q}\,
       =\, \frac{p_0 + \gamma p_z}{M}
\label{eq:y}
\end{eqnarray}
is the light-cone fraction of the nuclear momentum carried by the
interacting nucleon.
The coefficients ${\cal C}_{ij}$ are given by
\begin{eqnarray}
{\cal C}_{11}
&=& 1,							\nonumber\\
{\cal C}_{12}
&=& (\gamma^2-1) \frac{{\bm p}_\perp^2}{4 y^2 M^2},	\nonumber\\
{\cal C}_{22}
&=& \frac{1}{\gamma^2}
\left[
1 + \frac{(\gamma^2-1)}{2 y^2 M^2}
    \left( 2 p^2 + 3 {\bm p}_\perp^2 \right)
\right],						\label{eq:Cij}\\
{\cal C}_{LL}
&=& 1,							\nonumber\\
{\cal C}_{L2}
&=& (\gamma^2-1) \frac{{\bm p}_\perp^2}{y^2 M^2}.	\nonumber
\end{eqnarray}
Note that while in the $Q^2 \to \infty$ limit all the structure
functions are ``diagonal'', at finite $Q^2$ the transverse and
longitudinal structure functions $F_1^A$ and $F_L^A$ receive
contributions from both the nucleon's
$\widetilde{F}_1^N$ and $\widetilde{F}_2^N$
(or $\widetilde{F}_L^N$ and $\widetilde{F}_2^N$)
structure functions, whereas $F_2^A$ remains diagonal.

The $p^2$ dependence of the off-shell nucleon structure functions
$\widetilde{F}_i^N$ is, in itself, unphysical and must be interpreted
in the context of the $p^2$ dependence of the spectral function
${\cal F}_0^N$, such that only the total nuclear structure function
is physical.
Nevertheless, for a given nuclear wave function model which defines
the spectral function, one can extract the off-shell part of the
nucleon structure function phenomenologically.
For small nucleon virtualities, $|v| \ll 1$, where
	$v \equiv v(p^2) = (p^2 - M^2)/M^2$,
one can expand the off-shell nucleon structure functions
in a Taylor series around $p^2=M^2$,
\begin{eqnarray}
\widetilde{F}_i^N\left(x,Q^2,p^2\right)
&=& F_i^N\left(x,Q^2\right)\,
\Big( 1 + v(p^2)\, \delta f^N_i\left(x,Q^2\right) + {\cal O}(v^2)
\Big),\ \ \ \ i=1,2,L
\label{eq:Fexp1}      
\end{eqnarray}
where $F_i^N$ are the on-shell nucleon structure functions,
and the coefficient of the ${\cal O}(v)$ term is given by
\begin{eqnarray}
\delta f^N_i\left(x,Q^2\right)
&=& \left. \frac{\partial \log \widetilde{F}_i^N \left(x,Q^2,p^2\right)}
	       	{\partial \log v(p^2)}
    \right|_{v=0}.
\label{eq:Fexp2}
\end{eqnarray}
In earlier analyses, the off-shell function $\delta f^N_i$ was either
computed within simple spectator quark models~\cite{KMPW95, KP06, CJ11}
or extrated from empirical fits to nuclear structure function
data~\cite{KP06, CJ15, AKP17} assuming dependence on $x$ only.
Furthermore, typically it has been assumed that the same function
describes the off-shell modification of both the $F_1^N$ and $F_2^N$
(and $F_L^N$) structure functions,
  $\delta f^N_1 = \delta f^N_2 = \delta f^N_L \equiv \delta f^N$.
However, unlike in previous analyses which assumed also the isospin
independence of $\delta f^N$, here we allow the off-shell corrections
for the proton and neutron to differ,
  $\delta f^p_i \neq \delta f^n_i$.

The Taylor series expansion in Eq.~(\ref{eq:Fexp1}) allows the
derivation of simple expressions for the nuclear structure functions
in terms of on-shell and off-shell convolutions.
For the on-shell part, taking the first term in Eq.~(\ref{eq:Fexp1})
yields the familiar on-shell convolution approximation to the
nuclear structure functions~\cite{KP06, CJ15, Bickerstaff89},
\begin{subequations}
\label{eq:conv}
\begin{eqnarray}
xF_1^{A \, ({\rm on})}(x,Q^2)
&=& \sum_N \int dy
    \left[ f_{11}^N(y,\gamma)\,
	   \frac{x}{y} F_1^N\left(\frac{x}{y},Q^2\right)
	 + f_{12}^N(y,\gamma)\,
	   F_2^N\left(\frac{x}{y},Q^2\right)
    \right],\ \ \ \
\label{eq:conv1}					\\
F_2^{A \, ({\rm on})}(x,Q^2)
&=& \sum_N \int dy
    \left[ f_{22}^N(y,\gamma)\,
	   F_2^N\left(\frac{x}{y},Q^2\right)
    \right],						\\
\label{eq:conv2}
F_L^{A \, ({\rm on})}(x,Q^2)
&=& \sum_N \int dy
    \left[ f_{LL}^N(y,\gamma)\, F_L^N\left(\frac{x}{y},Q^2\right)
	 + f_{L2}^N(y,\gamma)\, F_2^N\left(\frac{x}{y},Q^2\right)
    \right],\ \ \ \
\label{eq:convL}
\end{eqnarray}
\end{subequations}
where the one-dimensional smearing functions are given by
\begin{eqnarray}
f_{ij}^N(y,\gamma)
&=& \int\!\frac{d^4{p}}{(2\pi)^4}
    {\cal F}_0^N\left(\eps,{\bm p}\right)
    \left( 1 + \frac{\gamma p_z}{M} \right)
    {\cal C}_{ij}\,
    \delta\left( y-1-\frac{\eps+\gamma p_z}{M} \right),
\label{eq:fij}
\end{eqnarray}
and the $y$ integrations in Eqs.~(\ref{eq:conv}) range
from $x$ to $M_A/M$.
Note that for $\gamma=1$ the diagonal functions $f_{ii}^N$
($i=1,2,L$) are normalized to unity, corresponding to the
spectral function normalization,
\begin{eqnarray}
\int_0^{M_A/M} dy\, f_{ii}^N(y,\gamma=1)
&=& \int\!\!\frac{d^4p}{(2\pi)^4}\,
    {\cal F}_0^N\left(\eps,{\bm p}\right)\
 =\ 1
\label{eq:fynorm}
\end{eqnarray}
for both protons and neutrons, $N=p, n$.
Generalizing Eqs.~(\ref{eq:conv}) to include the off-shell term
in Eq.~(\ref{eq:Fexp1}) proportional to $v$, one can write the
first-order off-shell contributions to the nuclear structure
functions as~\cite{Ehlers14}
\begin{subequations}
\label{eq:conv_off}
\begin{eqnarray}
xF_1^{A \, ({\rm off})}(x,Q^2)
&=& \sum_N \int\!dy
    \left[ \widetilde{f}_{11}^N(y,\gamma)\,
	   \frac{x}{y} F_1^N\left(\frac{x}{y},Q^2\right)
	 + \widetilde{f}_{12}^N(y,\gamma)\,
	   F_2^N\left(\frac{x}{y},Q^2\right)
    \right]					\nonumber\\
& & \hspace*{2cm} \times\
    \delta f^N\left(\frac{x}{y},Q^2\right),
\label{eq:yconv1}				\\
F_2^{A \, ({\rm off})}(x,Q^2)
&=& \sum_N \int\!dy
    \left[ \widetilde{f}_{22}^N(y,\gamma)\,
	   F_2^N\left(\frac{x}{y},Q^2\right)
    \right] \delta f^N\left(\frac{x}{y},Q^2\right),
\label{eq:yconv2}				\\
xF_L^{A \, ({\rm off})}(x,Q^2)
&=& \sum_N \int\!dy
    \left[ \widetilde{f}_{LL}^N(y,\gamma)\,
	   F_L^N\left(\frac{x}{y},Q^2\right)
	 + \widetilde{f}_{L2}^N(y,\gamma)\,
	   F_2^N\left(\frac{x}{y},Q^2\right)
    \right]					\nonumber\\
& & \hspace*{2cm} \times\
    \delta f^N\left(\frac{x}{y},Q^2\right),
\label{eq:yconvL}
\end{eqnarray}
\end{subequations}
where the off-shell smearing functions $\widetilde{f}_{ij}^N$
are defined by including the factor $v$ in the integrand of
Eq.~(\ref{eq:fij}),
\begin{eqnarray}
\widetilde{f}_{ij}^N(y,\gamma)
&=& \int\!\frac{d^4{p}}{(2\pi)^4}
    {\cal F}_0^N\left(\eps,{\bm p}\right)
    \left( 1 + \frac{\gamma p_z}{M} \right)
    {\cal C}_{ij}\, v(p^2)\,
    \delta\left( y-1-\frac{\eps+\gamma p_z}{M} \right).
\label{eq:fij_off}
\end{eqnarray}
The total nuclear structure functions are then given by the sum of
the on-shell and off-shell contributions,
\begin{eqnarray}
F_i^A(x,Q^2)
&=& F_i^{A\, \rm (on)}(x,Q^2)\,
+\, F_i^{A\, \rm (off)}(x,Q^2),\ \ \ \ i=1,2,L.
\label{eq:dis_onoff}
\end{eqnarray}
These results are general and valid for any bound system of $A$
nucleons.  With the above normalization for the smearing functions
(\ref{eq:fynorm}), the nuclear structure functions can be written
in terms of the proton and neutron contributions as
\begin{eqnarray}
F_i^A(x,Q^2)
&=& Z F_i^{p/A}(x,Q^2)\, +\, (A-Z) F_i^{n/A}(x,Q^2),\ \ \ \ i=1,2,L.
\label{eq:dis_pn}
\end{eqnarray}
In the next section we specialize to the case of $A=3$ nuclei.

\subsection{Smearing functions for $A=3$ nuclei}
\label{ssec:A3smear}

In this section we describe the proton and neutron spectral functions
for the case of three-body nuclei, which is the focus of the present
study, and illustrate the shapes and magnitudes of the on-shell and
off-shell smearing functions for specific models.
In general, the spectral function is defined to give the probability
distribution for finding a nucleon with momentum $\bm{p}$ and energy
$\eps$ in the nucleus $A$, summed over all possible configurations
of the residual $A-1$ system.
For the proton spectral function in $^3$He there are two contributions:
one from the bound $pn$ intermediate state corresponding to a deuteron,
with energy 
	$\eps = \eps_d - \eps_{^3{\rm He}}$,
where $\eps_d = -2.22$~MeV and $\eps_{^3{\rm He}} = -7.72$~MeV
are the deuteron and $^3$He binding energies, respectively,
and one from the $pn$ continuum scattering states, with off-shell
energy $\eps$,
\begin{eqnarray}
{\cal F}_0^p(\eps,{\bm p})
&=& {\cal F}_0^{p\, (d)}({\bm p})\
    \delta( \eps + \eps_{^3{\rm He}} - \eps_d )\
 +\ {\cal F}_0^{p\, ({\rm cont})}(\eps,{\bm p}).
\label{eq:proton_spec}
\end{eqnarray}
For the neutron, on the other hand, since there is no bound state of
two protons, the spectral function contains only a contribution from
the $pp$ continuum scattering state,
\begin{eqnarray}
{\cal F}_0^n(\eps,{\bm p})
&=& {\cal F}_0^{n\, ({\rm cont})}(\eps,{\bm p}).
\label{eq:neutron_spec}
\end{eqnarray}
Assuming isospin symmetry, the spectral functions for tritium, $^3$H,
can be obtained from those of $^3$He simply by interchanging the
proton and neutron.  As is well known, however, this underestimates
the triton binding energy of $\eps_{^3{\rm H}} = -8.482$~MeV, and
requires the addition of Coulomb interactions and charge-symmetry
breaking effects.

In practice, the spectral functions are typically evaluated in terms
of the (positive) separation energy $E$, defined as the energy
required to remove a single (on-shell) nucleon from the nucleus,
\begin{eqnarray}
E &=& M_{A-1} + M - M_A,
\label{eq:sepE}
\end{eqnarray}
where the mass of the spectator $A-1$ system is
\begin{eqnarray}
M_{A-1} &=& \sqrt{E_{A-1}^2 - {\bm p}_{A-1}^2},
\label{eq:MA-1}
\end{eqnarray}
with 
\begin{eqnarray}
E_{A-1} &=& M_A - p_0\ =\ M_A - M - \eps
\label{eq:EA-1}
\end{eqnarray}
the on-shell energy of the spectator system, and
${\bm p}_{A-1}^2 = {\bm p}^2$ in the rest frame of the nucleus.
Solving Eqs.~(\ref{eq:sepE}) and (\ref{eq:EA-1}), the energy
$\eps$ can be written in terms of the separation energy $E$ as
\begin{eqnarray}
\eps &=& M_A - M - \sqrt{(E + M_A - M)^2 + {\bm p}^2},
\label{eq:epsE}
\end{eqnarray}
which in the nonrelativistic limit is approximated as
\begin{eqnarray}
\eps &\approx& - E - \frac{{\bm p}^2}{2 (E+M_A-M)}.
\label{eq:epsNR}
\end{eqnarray}
For a nucleon at rest in the nucleus, ${\bm p}=0$, the energy
$\eps$ is then simply the negative of the separation energy,
\begin{eqnarray}
\eps({\bm p}=0) &=& - E.
\end{eqnarray}

The functions ${\cal F}_0^{p\, (d)}({\bm p})$ and
${\cal F}_0^{p, n\, ({\rm cont})}$ can be determined by solving
the three-body bound state problem using one of several methods.
Bissey {\it et al.}~\cite{Bissey01} solved the Faddeev equation
using a separable approximation to the two-body nucleon--nucleon
Paris potential~\cite{Haidenbauer84}, as well as the unitary pole
approximation~\cite{Saito95} to the Reid soft core (RSC) $NN$
potential~\cite{RSC68}, and the Yamaguchi potential~\cite{Yamaguchi54}
with 7\% mixing between $^3S_1$ and $^3D_1$ waves.
The resulting smearing functions were used in the analysis of $^3$He
and $^3$H structure functions in Refs.~\cite{Afnan00, Afnan03}.
Schulze and Sauer (SS)~\cite{SS} also solved the Faddeev equation
for 18 channels using the Paris $NN$ potential for the ground state
$^3$He wave function~\cite{Stadler91}, and projecting onto the
deuteron and continuum scattering states.

In contrast, Ciofi degli Atti {\it et al.} pioneered~\cite{Ciofi80,
Ciofi84} the use of the variational method using harmonic
oscillator wave functions and the RSC $NN$ interaction.
Kievsky~{\it et~al.} (KPSV)~\cite{KPSV} extended this approach,
making use of a pair-correlated hyperspherical harmonic basis
\cite{Kievsky93} with the AV18 $NN$ potential, including a Coulomb
interaction between protons and the Urbana IX three-body force.
The KPSV and SS spectral functions were used in the analyses of
spin-dependent $^3$He structure functions in Refs.~\cite{KM08,
Ethier13}, and we will use these in the present work.
Table~\ref{tab:specfunc} summarizes the average nucleon off-shell
energy $\eps$ and kinetic energy $\langle{\bm p}^2\rangle/2M$
for the KPSV~\cite{KPSV} and SS~\cite{SS} models.

\begin{table}
\caption{Average nucleon energy $\eps$ and kinetic energy
	$\langle{\bm p}^2\rangle/2M$ (in units of MeV) in $^3$He
	and $^3$H nuclei, for the KPSV~\cite{KPSV} and SS~\cite{SS}
	models of the nuclear spectral functions.}
\centering
\begin{tabular}{{
>{\centering\arraybackslash}m{1in} 
>{\centering\arraybackslash}m{1in}
>{\centering\arraybackslash}m{1in}
>{\centering\arraybackslash}m{1in}}}
  model & nucleus & $\langle{\eps}\rangle$
		  & $\langle{\bm p}^2\rangle/2M$	\\ \hline
  KPSV  & $^3$He  & $-64.28$ & 48.85	\\
        & $^3$H   & $-66.56$ & 48.84	\\ \hline
  SS    & $^3$He  & $-53.66$ & 38.45	\\
        & $^3$H   & $-55.94$ & 38.44	\\ \hline
\end{tabular}
\label{tab:specfunc}
\end{table}

\begin{figure}[t]
\includegraphics[width=16cm]{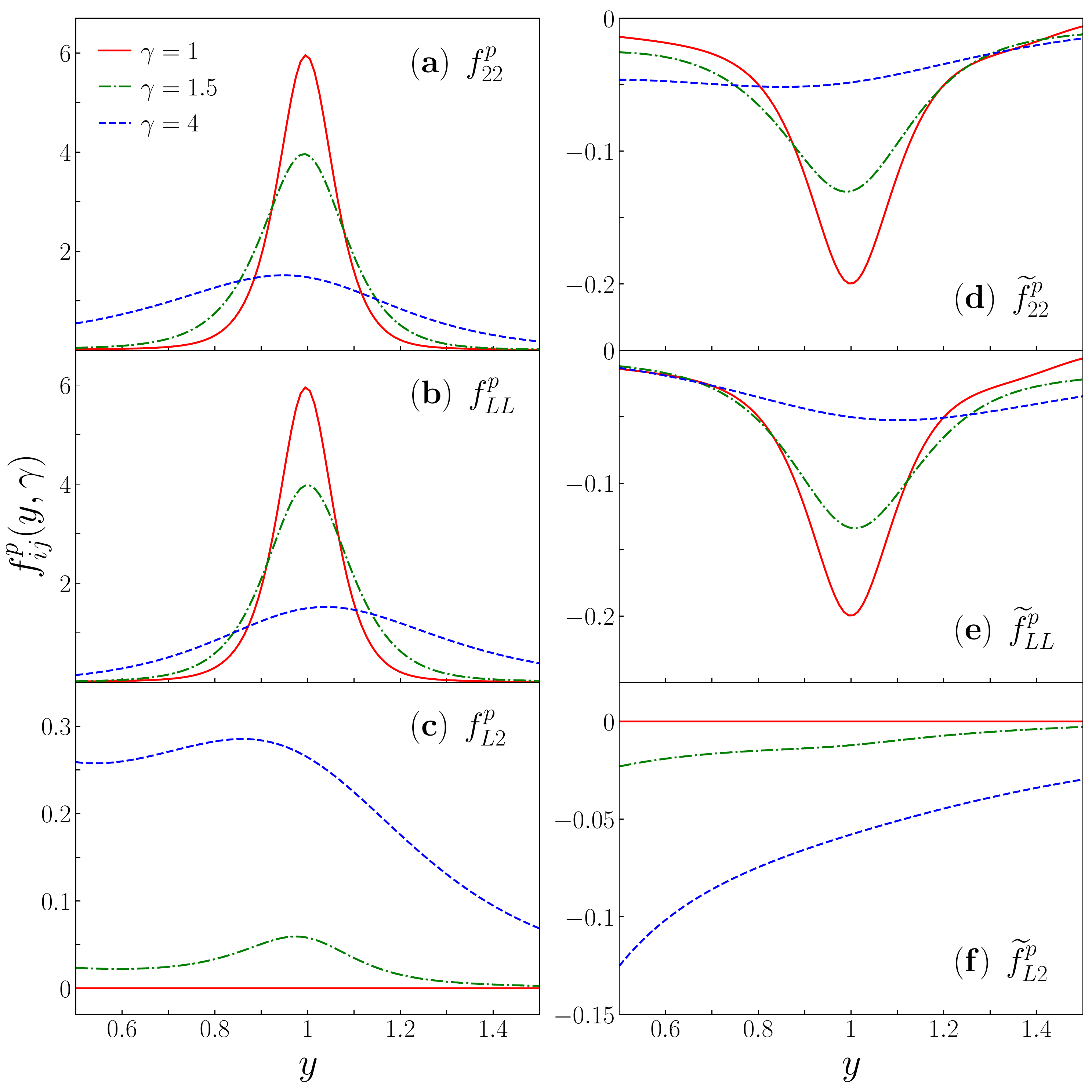}
\vspace*{-0.5cm}
\caption{Proton smearing functions in $^3$He, $f_{ij}^p(y,\gamma)$,
	$i=2,L$, for the on-shell [(a)--(c)] and off-shell [(d)--(f)]
	distributions, computed from the KPSV spectral
	function~\cite{KPSV} for $\gamma=1$ (red solid curves),
	1.5 (green dot-dashed curves) and 4 (blue dashed curves).}
\label{fig:fyp}
\end{figure}

\begin{figure}[t]
\includegraphics[width=16cm]{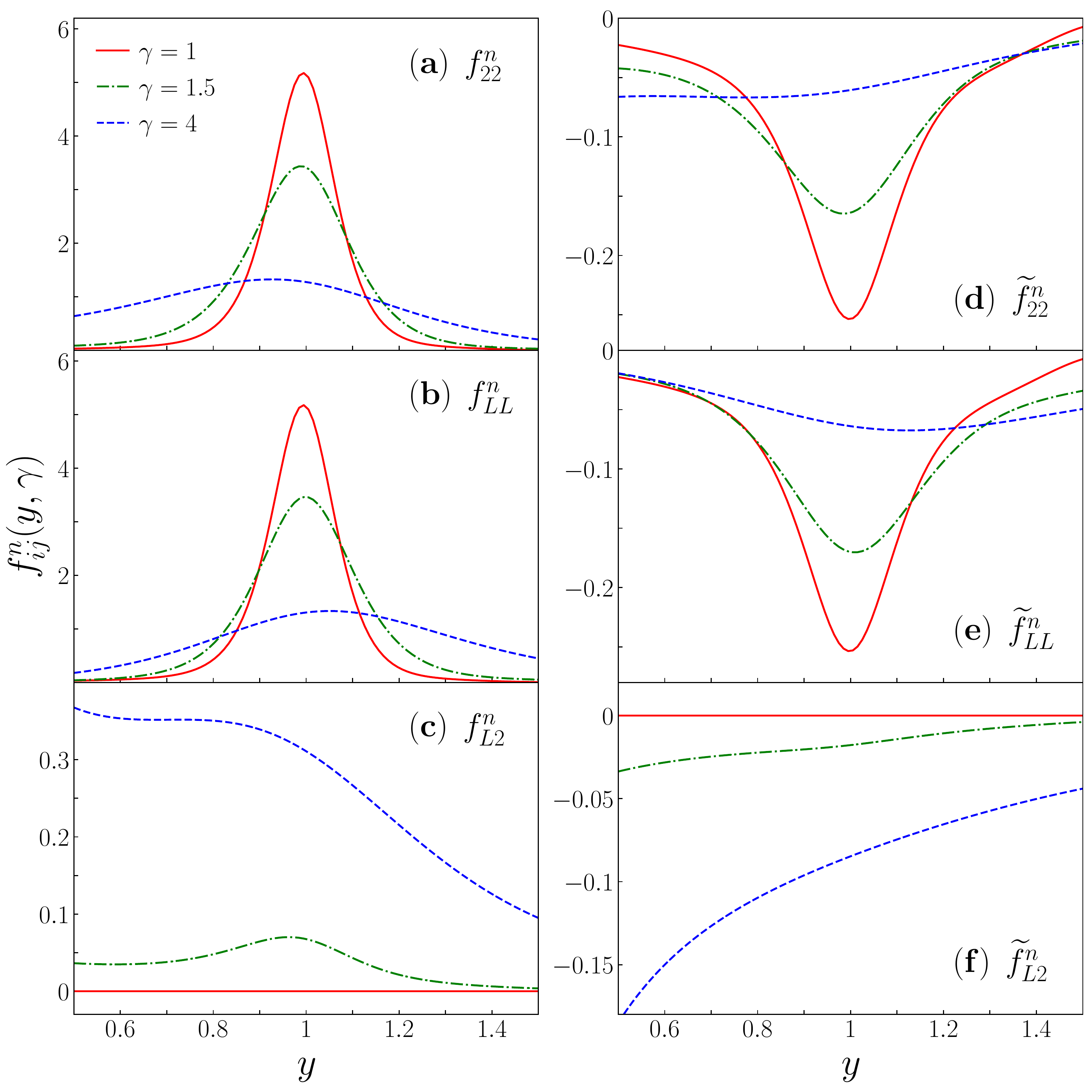}
\vspace*{-0.5cm}
\caption{Neutron smearing functions in $^3$He, $f_{ij}^n(y,\gamma)$,
	$i=2,L$, for the on-shell [(a)--(c)] and off-shell [(d)--(f)]
	distributions, computed from the KPSV spectral
	function~\cite{KPSV} for $\gamma=1$ (red solid curves),
	1.5 (green dot-dashed curves) and 4 (blue dashed curves).}
\label{fig:fyn}
\end{figure}

\begin{figure}[t]
\includegraphics[width=16cm]{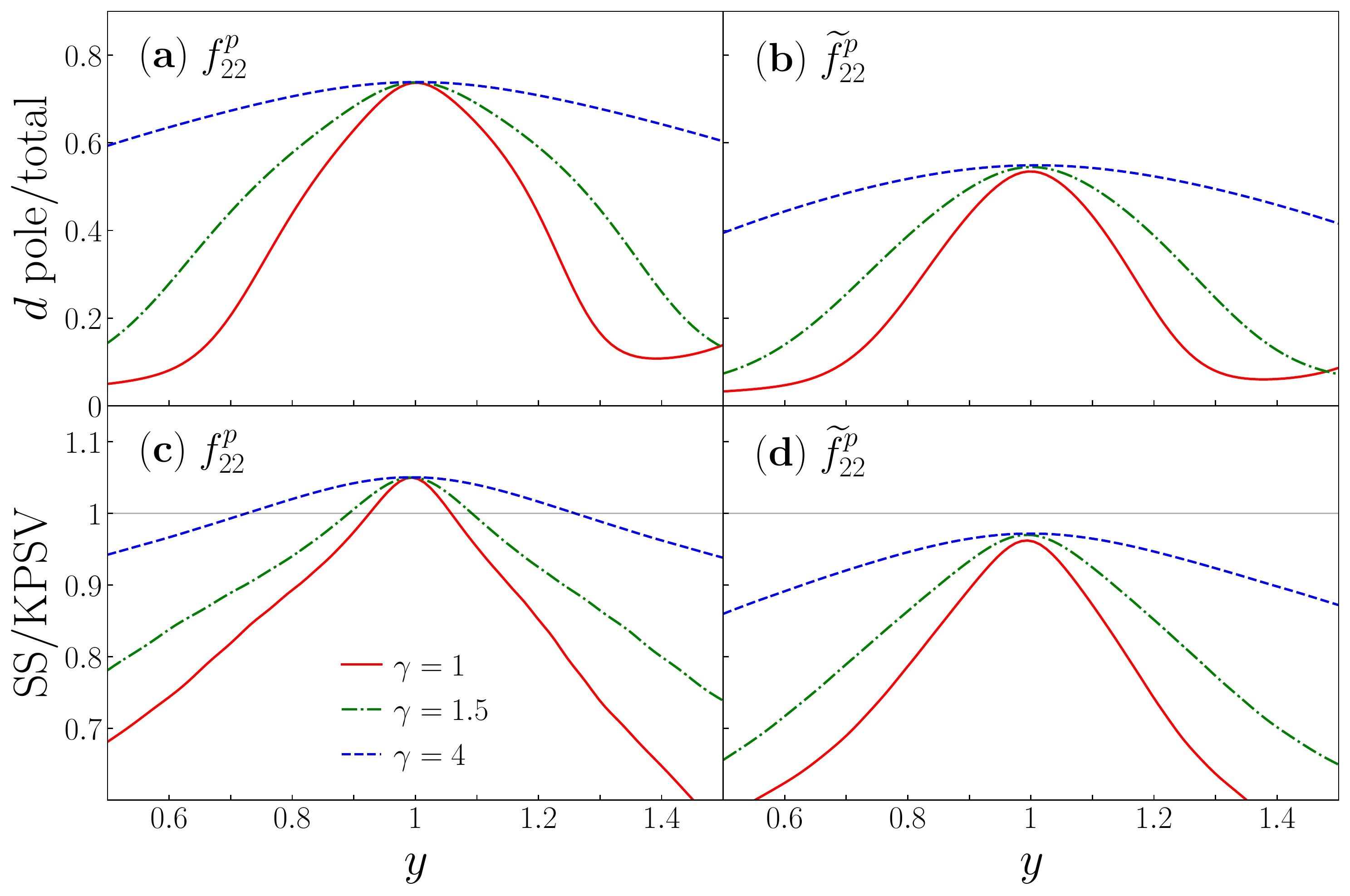}
\vspace*{-0.5cm}
\caption{Ratio of deuteron pole contribution to the total smearing
	function for (a) the proton on-shell $f_{22}^p$ function,
	and (b) proton off-shell $\widetilde{f}_{22}^p$ function,
	for different values of $\gamma$.
	The ratio of the total proton smearing functions for the
	SS~\cite{SS} and KPSV~\cite{KPSV} spectral functions is
	given in (c) and (d) for the on-shell and off-shell functions,
	respectively.}
\label{fig:fyrat}
\end{figure}

The on-shell smearing functions $f_{ij}^N$ for the proton and neutron
in $^3$He, as well as the off-shell functions $\widetilde{f}_{ij}^N$,
are illustrated in Figs.~\ref{fig:fyp} and \ref{fig:fyn}, respectively,
for the KPSV model, at several values of the parameter $\gamma$.
The diagonal functions $f_{22}^N$ and $f_{LL}^N$ are steeply peaked
around $y=1$, but become broader with increasing $\gamma$.
This has the effect of introducing stronger nuclear smearing at low
values of $Q^2$ and at large $x$, where the cross sections are small,
than at lower $x$ where they are larger.
Note that for $\gamma=1$ the $f_{22}^N$ and $f_{LL}^N$
distributions are identical, but differ for $\gamma > 1$.
The non-diagonal functions $f_{L2}^N$ vanish identically for
$\gamma=1$, but rise to $\sim 20\%$ of the diagonal functions
at $y=1$ for $\gamma=4$.

For the off-shell smearing functions in Figs.~\ref{fig:fyp} and
\ref{fig:fyn}, because of the factor $v$ ($< 0$) in the integrand
of Eq.~(\ref{eq:fij_off}), these are negative for both the proton
and neutron.
For $\gamma=1$, the off-shell functions $\widetilde{f}_{22}^N$
and $\widetilde{f}_{LL}^N$ are identical, with a magnitude of
$\approx 3\%$ of their on-shell counterparts at the peak
$y \sim 1$ for the proton and $\approx 5\%$ for the neutron.
As for the on-shell functions, the off-shell distributions become
broader with increasing $\gamma$, approximately tracking the
$\gamma$ dependence of the on-shell distributions.

The slightly narrower peak for the proton function in Fig.~\ref{fig:fyp}
compared with the neutron in Fig.~\ref{fig:fyn} reflects the presence
of the bound deuteron spectator contribution in the former but not
in the latter.
In fact, the deuteron bound state component amounts to around 2/3
of the strength of the proton on-shell smearing function, with the
continuum contribution accounting for $\sim 1/3$.
This is illustrated in Fig.~\ref{fig:fyrat}(a) and (b),
where the deuteron contribution is shown relative to the total for
the proton on-shell $f_{22}^p$ and off-shell $\widetilde{f}_{22}^p$
functions, respectively.
For the proton off-shell function, the fraction at the
$y \approx 1$ peak is closer to 1/2.
Away from the peak, the deuteron pole fractions decrease rapidly
for $\gamma=1$, but remain broader for larger $\gamma$.
The results for the $f_{LL}^p$ and $\widetilde{f}_{LL}^p$
functions are very similar to those in Fig.~\ref{fig:fyrat},
as are the ratios for the neutron.

The dependence of the smearing functions on the choice of model for the
$A=3$ wave function is illustrated in Fig.~\ref{fig:fyrat}(c) and (d)
as a ratio of SS~\cite{SS} to KPSV~\cite{KPSV} spectral functions for
the proton $f_{22}^p$ and $\widetilde{f}_{22}^p$ distributions.
The on-shell smearing function for the SS model is slightly
narrower around $y \approx 1$, with an $\approx 5\%$ higher peak,
which is compensated by lower distributions away from the peak.
For the off-shell function, the SS model distribution is
$\approx 5\%$ lower than for the KPSV model, with a similar
behavior away from the peak.
These results illustrate an interesing compensation for
the differences between the on-shell smearing functions
around $y=1$ and the off-shell functions for the two models.

With these distributions, one can now proceed to compute the
nuclear structure functions $F_i^A$ for $A=$ $^3$He and $^3$H,
which will be the subject of the remaining sections.

\section{Quasielastic Scattering}
\label{sec:qe}

The veracity of any calculation of nuclear structure functions
depends, within the convolution framework of Eqs.~(\ref{eq:conv})
and (\ref{eq:conv_off}), on the reliability of the smearing
functions $f_{ij}^N(y,\gamma)$ that characterize the distribution
of nucleons in the nucleus.
One of best testing grounds for models of the smearing functions
is QE electron--nucleus scattering, where an electron scatters
elastically from a proton or neutron bound in the nucleus.
Whereas for inelastic scattering from the bound nucleon the
light-cone distributions are convoluted with a nontrivial
$x$ distribution in the inelastic $F_i^N$ structure functions,
for QE scattering the cross section and structure functions are
given directly by products of $f_{ij}^N$ and $Q^2$-dependent
elastic nucleon form factors.
It is important, therefore, to establish the range of kinematics
whereby the inclusive cross sections can be described in terms
of the {\it same} smearing functions in both the deep-inelastic
and QE regions.
After providing the basic formulas for the QE contributions to the
nuclear structure functions, in this section we compare the results
for the $^3$He cross sections with precision QE data from SLAC and
Jefferson Lab in the vicinity of $x \approx 1$.

\subsection{Quasielastic structure functions}

The matrix elements of the electromagnetic current operator $J^\mu$
between on-shell nucleon states are usually parametrized in terms
of the nucleon's Dirac $F_{1N}$ and Pauli $F_{2N}$ form factors,
\begin{eqnarray}
\langle N(p+q) | J^\mu | N(p) \rangle
&=& \bar{u}(p+q)
    \left[ \gamma^\mu\, F_{1N}(Q^2)\
	+\ i \sigma^{\mu\nu} q_\nu \frac{F_{2N}(Q^2)}{2M}
    \right]
    u(p).
\label{eq:Jcc2}
\end{eqnarray}
Using the Dirac equation, and defining the Sachs electric $G_{EN}$
and magnetic $G_{MN}$ form factors such that
\begin{subequations}
\label{eq:ffs}
\begin{eqnarray}
F_{1N}(Q^2)
&=& \frac{1}{1+\tau}
    \left[ G_{EN}(Q^2) + \tau G_{MN}(Q^2) \right],
\label{eq:ff1}					\\
F_{2N}(Q^2)
&=& \frac{1}{1+\tau} 
    \left[ G_{MN}(Q^2) - G_{EN}(Q^2) \right],
\label{eq:ff2}
\end{eqnarray}
\end{subequations}
where $\tau = 4M^2/Q^2$, the matrix element can be equivalently
written as
\begin{eqnarray}
\langle N(p+q) | J^\mu | N(p) \rangle
&=& \bar{u}(p+q)
    \left[ \gamma^\mu\, G_{MN}(Q^2)\
	-\ (2 p^\mu + q^\mu) \frac{F_{2N}(Q^2)}{2M}
    \right]
    u(p).
\label{eq:Jcc1}
\end{eqnarray}
The contributions to the nucleon elastic structure functions $F_i^N$
are then given by products of the form factors multiplied by an energy
conserving $\delta$ function at $x=1$,
\begin{subequations}
\label{eq:F12el}
\begin{eqnarray}
F_1^{N (\rm el)}(x,Q^2)
&=& \left[ \frac{1}{2} G_{MN}^2(Q^2) \right]
    \delta(1-x),
\label{eq:F1el}					\\
F_2^{N (\rm el)}(x,Q^2)
&=& \left[ \frac{G_{EN}^2(Q^2) + \tau G_{MN}^2(Q^2)}{1+\tau} \right]
    \delta(1-x),
\label{eq:F2el}
\end{eqnarray}
\end{subequations}
where we have used the on-shell relation
\begin{eqnarray}
Q^2\, \delta\big( (p+q)^2-M^2 \big)
&=& 2 p\cdot q\, \delta\big( (p+q)^2-M^2 \big)\,
 =\, \delta(1-x).
\label{eq:delta1-x}
\end{eqnarray}
The elastic contribution to the longitudinal structure function,
$F_L^{N (\rm el)}$, can then be computed from Eqs.~(\ref{eq:F12el})
using the relation
$F_L^{N (\rm el)}(x,Q^2)
= (1+1/\tau) F_2^{N (\rm el)}(x,Q^2)
- 2 F_1^{N (\rm el)}(x,Q^2)$.
Putting these results together, the QE nuclear structure functions
can be written in terms of the nucleon Sachs form factors as
\begin{subequations}
\label{eq:F12QE}
\begin{eqnarray}
x F_1^{A (\rm QE)}(x,Q^2)
&=& \sum_N
\left\{
    \frac{1}{2} x f_{11}^N(x,\gamma)\,
    G_{M N}^2(Q^2)\
 +\ x f_{12}^N(x,\gamma)
    \left[ \frac{G_{E N}^2(Q^2) + \tau G_{M N}^2(Q^2)}{1+\tau}
    \right]
\right\},
\label{eq:F1QE}				\nonumber\\
& &					\\
F_2^{A (\rm QE)}(x,Q^2)
&=& \sum_N
    x f_{22}^N(x,\gamma)
    \left[ \frac{G_{E N}^2(Q^2) + \tau G_{M N}^2(Q^2)}{1+\tau}
    \right].
\label{eq:F2QE}
\end{eqnarray}
\end{subequations}
%

\subsection{Off-shell nucleons}

Generalizing to the case where the struck nucleon is bound inside
a nucleus, and hence off its mass-shell, $p^2 \not= M^2$, one can
write the kinematic constraint for elastic scattering to an on-shell
nucleon final state as
    $2 p \cdot q = Q^2 + M^2 - p^2 = Q^2/(x/y)$,
where $y$ is defined in Eq.~(\ref{eq:y}).
In this case the $\delta$ function in Eq.~(\ref{eq:delta1-x}) can be
written as
\begin{eqnarray}
Q^2\, \delta\big((p+q)^2 - M^2\big)
&=& \frac{x}{y}\
    \delta\Big(1 - \kappa(p^2)\, \frac{x}{y}\Big),
\label{eq:deltafn-off}
\end{eqnarray}
where $\kappa(p^2) = 1 - v(p^2)\, M^2/Q^2$ parametrizes the
kinematical effects of the off-shell correction.
The generalization of the current operator to off-shell is not unique,
and in the literature one encounters several prescriptions for this.
The most common ones are the ``cc1'' and ``cc2'' prescription of
De~Forest~\cite{DeForest83}, which correspond to generalizing the
currents in Eqs.~(\ref{eq:Jcc1}) and (\ref{eq:Jcc2}), respectively,
to the off-shell region.
The elastic structure functions for the off-shell nucleon are then
given by 
\begin{subequations}
\label{eq:qe_off2}
\begin{eqnarray}
\widetilde{F}_1^{N (\rm el)}\Big(\frac{x}{y},Q^2,p^2\Big)
&=& \left[ \frac{G_{MN}^2}{2}
	   \left( 1 + \frac{v M^2}{Q^2} \right)
    \right]
    \frac{x}{y} \delta\Big(1-\kappa(p^2)\, \frac{x}{y}\Big),\\
\widetilde{F}_2^{N (\rm el)}\Big(\frac{x}{y},Q^2,p^2\Big)
&=& \left[ \frac{G_{EN}^2 + \tau G_{MN}^2}{1+\tau}
	 + v\, \frac{(G_{MN}-G_{EN})^2}{4 (1+\tau)^2}
    \right]
    \delta\Big(1-\kappa(p^2)\, \frac{x}{y}\Big),
\end{eqnarray}
\end{subequations}
and
\begin{subequations}
\label{eq:qe_off1}
\begin{eqnarray}
\widetilde{F}_1^{N (\rm el)}\Big(\frac{x}{y},Q^2,p^2\Big)
&=& \left[
      \frac{G_{MN}^2}{2}
    + \frac{v M^2}{2 Q^2}
      \left( \frac{G_{EN}^2+\tau G_{MN}^2}{1+\tau}
	   + v\, \frac{(G_{MN}-G_{EN})^2}{4 (1+\tau)^2}
      \right)
    \right]						\nonumber\\
& & \times\,
    \frac{x}{y} \delta\Big(1-\kappa(p^2)\, \frac{x}{y}\Big),\\
\widetilde{F}_2^{N (\rm el)}\Big(\frac{x}{y},Q^2,p^2\Big)
&=& \left[ \frac{G_{EN}^2 + \tau G_{MN}^2}{1+\tau}
    \right]
    \delta\Big(1-\kappa(p^2)\, \frac{x}{y}\Big),
\end{eqnarray}
\end{subequations}
for the ``cc1'' and ``cc2'' cases, respectively.
Assuming the $G_E$ and $G_M$ form factors themselves remain
functions of $Q^2$ only, the off-shell corrections to the
on-shell elastic structure functions in Eqs.~(\ref{eq:F12el})
involve terms that are of order $v$ and $v^2$, in addition to
the modified $\delta$~function.  In each case the off-shell
corrections vanish in the $Q^2 \to \infty$ limit.
In terms of the elastic off-shell functions, the total QE structure
functions can be computed by substituting Eqs.~(\ref{eq:qe_off2})
or (\ref{eq:qe_off1}) into Eqs.~(\ref{eq:F1F2_gen}), and using
the $\delta$ function in (\ref{eq:deltafn-off}) to eliminate the
$dy$ integration in
	$d^4p \to dy\, d|\bm{p}|\, d\eps$,
so that the QE structure functions are computed as integrals over
the variables $|\bm{p}|$ and $\eps$.
Alternatively, one can use the $\delta$ function to eliminate the
$|\bm{p}|$ or $p^2$ integration, leaving integrations over $\eps$
and $y$.

In the following we discuss the off-shell corrections numerically,
and compare the WBA predictions for the QE cross sections with
experimental measurements of the inclusive cross sections in the
QE region at $x \sim 1$.

\subsection{Comparison with quasielastic $^3$He data}

A number of experiments have been performed scattering electrons
from $A=3$ nuclei in the QE region, over a range of energies and
scattering angles.
A convenient summary of the experimental data is provided in the
Quasielastic Electron--Nucleus Scattering Archive~\cite{DayQE}.
The most relevant of these for the present analysis are data
from experiments at SLAC~\cite{Day:1979bx, Meziani:1992xr}
and Jefferson Lab~\cite{Fomin:2010ei}.

The early SLAC data from Ref.~\cite{Day:1979bx} were taken for
incident electron energies between 3 and 15~GeV at $\theta=8^\circ$
scattering angle, corresponding to momentum transfers of up to
$\approx 1.4$~GeV.
Measurements from the subsequent NE9 experiment~\cite{Meziani:1992xr}
were taken at electron energies between 0.9 and 4.3~GeV, and
scattering angles of $15^\circ$ and $85^\circ$.
Both the transverse and longitudinal structure functions were
extracted using the Rosenbluth separation technique at a 3-momentum
transfer of $\approx 1$~GeV, and the latter was used to test the
Coulomb sum rule.
More recently, high precision data from the Jefferson Lab experiment
E02-019 were collected using a 5.766~GeV electron beam on various
nuclear targets, including $^3$He, primarily to study ``super-fast''
quarks at $x > 1$~\cite{Fomin:2010ei}.  QE data were taken
at scattering angles between $18^\circ$ and $50^\circ$,
corresponding to values of the four-momentum exchange squared
of $2 \lesssim Q^2 \lesssim 9$~GeV$^2$.

Data from lower energy experiments from Bates \cite{Dow:1988rk} and
Saclay \cite{Marchand:1985us}, on $^3$He as well as $^3$H targets,
are not included in our analysis, which focuses on the region of
validity of the nuclear impulse approximation, corresponding to
intermediate $Q^2$ values from $\sim 1$~GeV$^2$ to a few~GeV$^2$.
At very large values of $x \gg 1$, contributions from processes
involving nucleons that no longer retain their clear identity as
nonoverlapping bound states of quarks, as well as multi-nucleon
effects requiring nuclear quark degrees of freedom, are expected
to become more important.
At very low $Q^2$ values, $Q^2 \ll 1$~GeV$^2$, coherent effects
and meson exchange corrections, as well as rescattering, are known
to play a greater role.
At higher $Q^2$ values, $Q^2 \gg 1$~GeV$^2$, identification of the
QE contribution underneath the rising inelastic background becomes
increasingly more difficult and model dependent.
In the $Q^2 \sim 1$~GeV$^2$ to several~GeV$^2$ range, where the
$x \sim 1$ region should still be dominated by single-nucleon
QE scattering, one can explore the efficacy and limitations of
an incoherent impulse approximation description in terms of the
nucleon smearing functions of Sec.~\ref{sec:theory}.

\begin{figure}[t]
\includegraphics[width=16cm]{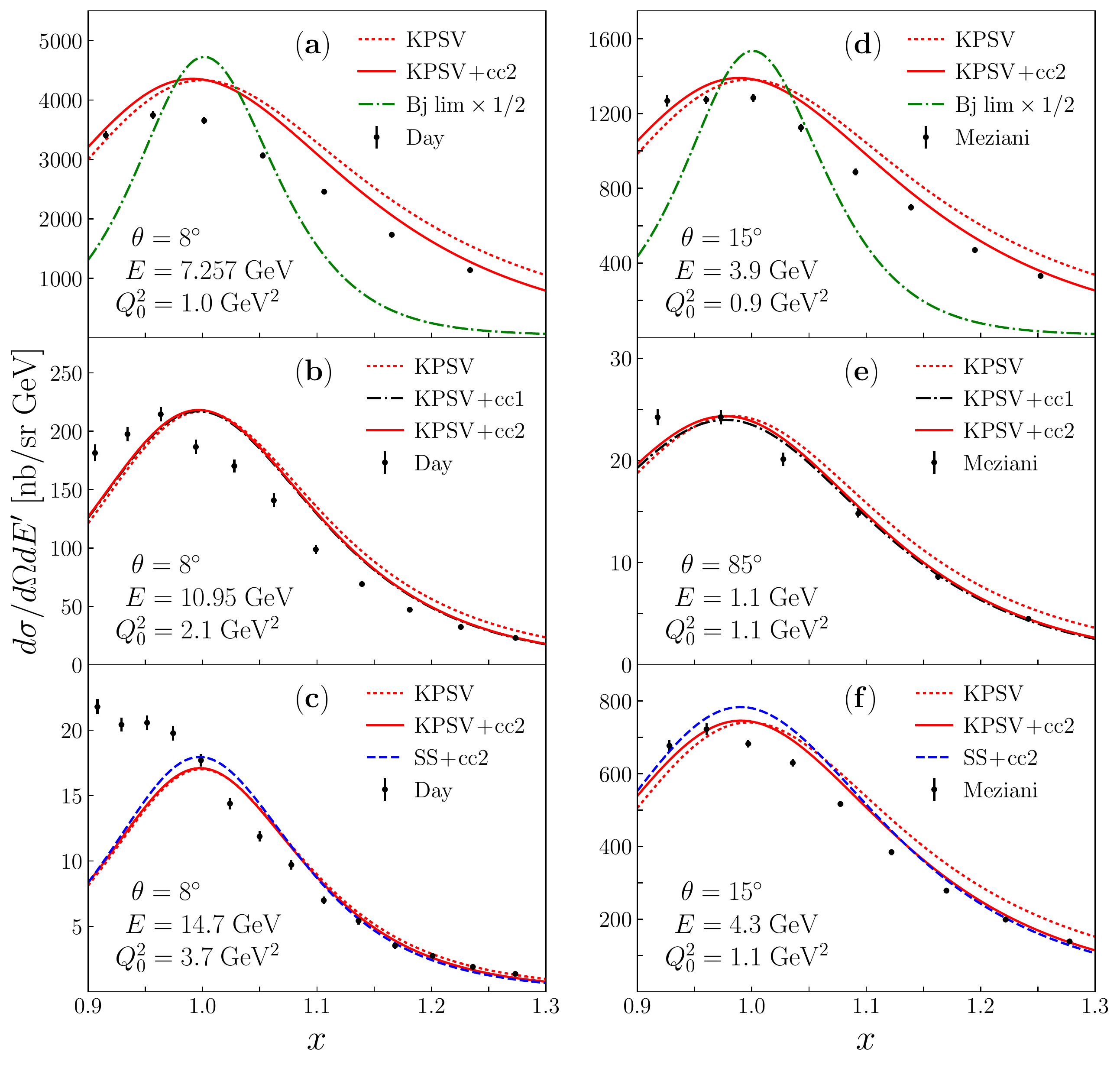}
\vspace*{-0.5cm}
\caption{QE electron--$^3$He cross section as a function
	of $x$ for fixed incident electron energies $E$ and
	scattering angles $\theta$, with $Q_0^2$ the value
	of the momentum transfer squared at $x=1$.
	Data from the early SLAC experiment by
	Day {\it et al.}~\cite{Day:1979bx} (a)--(c)
	and the subsequent NE9 experiment by
	Meziani {\it et al.}~\cite{Meziani:1992xr} (d)--(f)
	are compared with the WBA calculation using the KPSV and SS
	spectral functions, and the ``cc1'' and ``cc2'' off-shell
	prescriptions, as well as a calculation using smearing
	functions at $\gamma=1$ [(a) and (d)], scaled by a
	factor 1/2 for clarity).}
\label{fig:QE_Day_Mez}
\end{figure}

In Fig.~\ref{fig:QE_Day_Mez} the QE data from the SLAC
experiments~\cite{Day:1979bx, Meziani:1992xr} are compared with the
cross sections computed from the smearing functions in the WBA model.
The data include both forward scattering angles
[Fig.~\ref{fig:QE_Day_Mez}(a)--(d), (f)], as well as sideways
scattering [Fig.~\ref{fig:QE_Day_Mez}(e)], with the value of
$Q^2$ at $x=1$ (labeled by $Q_0^2$) ranging from
$\approx 1$~GeV$^2$ to $\lesssim 4$~GeV$^2$.
For the elastic structure function we use the parametrizations
of the electric and magnetic form factors of the proton and neutron
from Kelly~\cite{Kelly04}.
Experience from previous analyses of QE scattering from the
deuteron~\cite{EthierQE} shows that use of other parametrizations,
{\it e.g.}, from Refs.~\cite{AMT07, Venkat11, Bosted95}, has little
($\lesssim$~few~\%) effect on the cross sections at the relavant
kinematics.

As a baseline for the calculation, the KPSV~\cite{KPSV} model is
used for the $^3$He spectral function, and the results with and
without the off-shell corrections are compared.
As Fig.~\ref{fig:QE_Day_Mez} illustrates, the effect of the off-shell
corrections is a softening of the momentum distribution, which shifts
the peak in the cross section to slightly lower values of $x$,
and improves the overall agreement with the data.
The difference between the off-shell corrections computed using
the two prescription (``cc1'' or ``cc2'') is very small, and,
as expected from Eqs.~(\ref{eq:deltafn-off})--(\ref{eq:qe_off1}),
the off-shell effects become less prominent with increasing $Q^2$.

The importance of the $Q^2$ dependence is illustrated more strikingly
in Fig.~\ref{fig:QE_Day_Mez}(a) and (d), which compares the calculation
using exact kinematics with that taking smearing functions at
$\gamma=1$, as often done in deep-inelastic scattering applications
at high $Q^2$.
The result with $\gamma=1$ gives a significantly narrower distribution
around $x=1$, and a peak that is $\sim 2$ times larger than the data
indicate.
In contrast, the results with the finite-$Q^2$ kinematics correctly
implemented is in significantly better agreement with the data.
The dependence of the results on the $^3$He spectral function is
very mild, as the comparison with the SS spectral function~\cite{SS}
in Fig.~\ref{fig:QE_Day_Mez}(c) and (f) shows, with the SS results
giving a slightly narrower distribution around the QE peak compared
with the KPSV spectral function~\cite{KPSV}.

Overall, the qualitative features of the data versus theory
comparisons are similar for the Day {\it et al.}~\cite{Day:1979bx}
and Meziani~{\it et al.}~\cite{Meziani:1992xr} data, with the
agreement being somewhat better for the more recent data
set~\cite{Meziani:1992xr}.
The similar kinematics of the two experiments, in particular for
forward scattering angles at $Q^2 \sim 1$~GeV$^2$, raise the
question of whether there may be a systematic underestimate
in the Day~{\it et al.}~\cite{Day:1979bx} data in this region.

\begin{figure}[t]
\includegraphics[width=16cm]{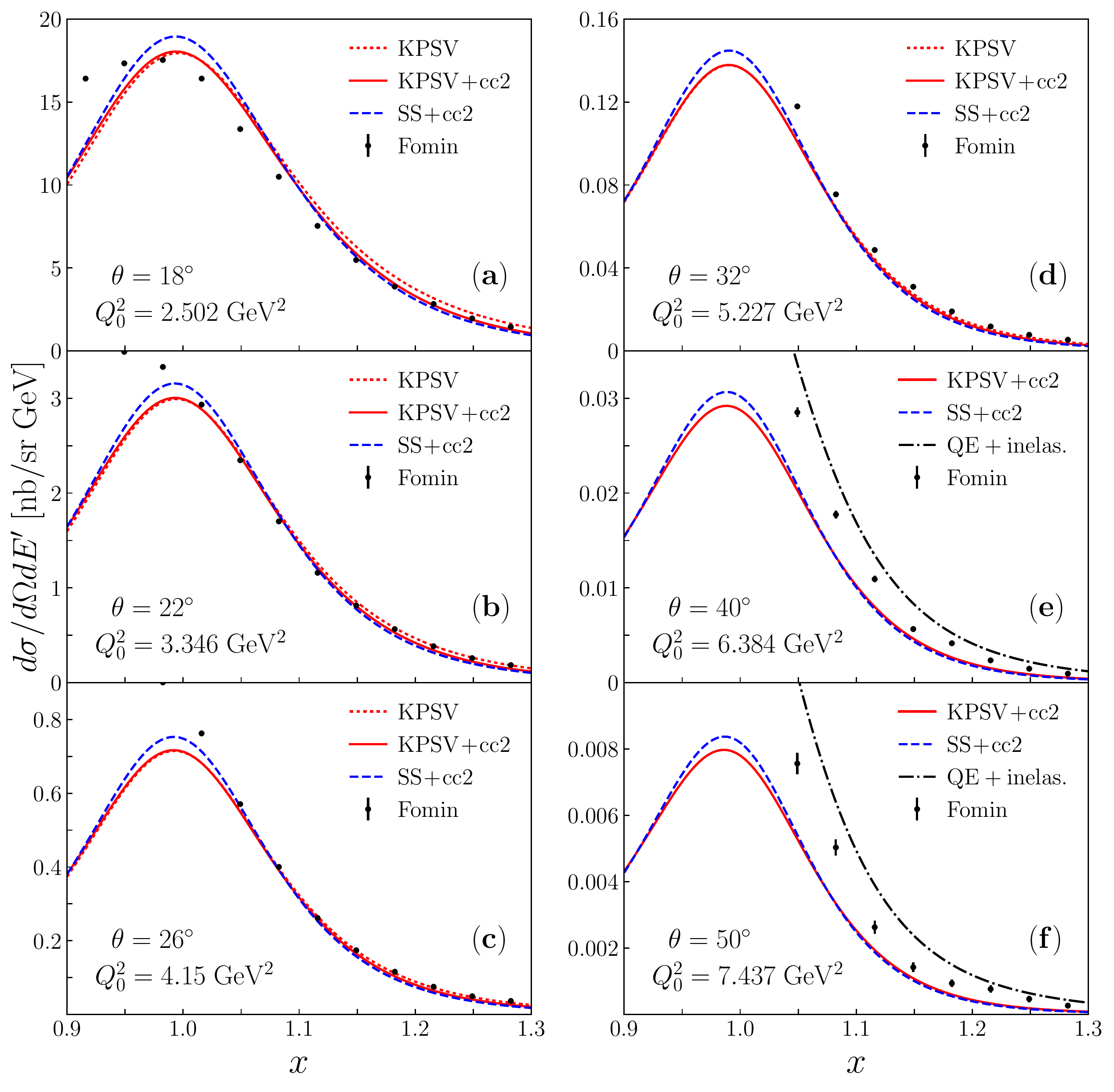}
\vspace*{-0.5cm}
\caption{As in Fig.~\ref{fig:QE_Day_Mez}, but for the Jefferson Lab
	QE data from Fomin {\it et al.}~\cite{Fomin:2010ei} at
	$E = 5.766$~GeV.
	Cross sections which include inelastic contributions are
	illustrated in panels (e) and (f) [black dot-dashed curves].}
\label{fig:QE_Fomin}
\end{figure}

The most recent QE data from Jefferson Lab experiment
E02-019~\cite{Fomin:2010ei} are shown in Fig.~\ref{fig:QE_Fomin},
for a fixed electron energy $E=5.766$~GeV and scattering angles
from $\theta=18^\circ$ to $50^\circ$.  This corresponds to slightly
larger $Q^2$ values at the QE peak than for the SLAC data in
Fig.~\ref{fig:QE_Day_Mez}, ranging from $Q_0^2 \approx 2.5$~GeV$^2$
to $\approx 7.4$~GeV$^2$.
At these higher $Q^2$ values the effects of the off-shell
corrections are relatively small, and for the spectra with
$Q_0^2 \gtrsim 4$~GeV$^2$ the full cross sections are almost
indistinguishable from the on-shell only contributions.

What is rather more important at the higher $Q^2$ values are
the effects of the inelastic background.  These are illustrated
in Fig.~\ref{fig:QE_Fomin}(e) at $Q_0^2 \approx 6.3$~GeV$^2$
and Fig.~\ref{fig:QE_Fomin}(f) at $Q_0^2 \approx 7.4$~GeV$^2$,
using the nonresonant background part of the Christy--Bosted
nucleon structure function parametrization~\cite{CB10}.
For the higher $Q_0^2$ case in particular, the inelastic cross section
is quite large --- more than half of the total at the QE peak.

Since the inelastic contribution in the present work is not fitted,
but simply ported from a previous analysis~\cite{CB10}, and since the
separation of the total cross section into resonance and nonresonant
background contributions is not unique, achieving quantitative agreement
of the QE plus inelastic sum with the data is not the primary goal.
The point to be noted in this comparison is the relative magnitude of 
the inelastic component compared with the nucleon elastic, and the 
difficulty in determining the QE piece unambiguously at high values of 
$Q^2$, $Q_0^2 \gtrsim 4-5$~GeV$^2$.

The dependence on the model $^3$He spectral function is similar
to that in Fig.~\ref{fig:QE_Day_Mez}, with the SS model~\cite{SS}
giving a slightly higher cross section at $x \approx 1$, with
marginally softer distributions away from the QE peak.
Overall, the agreement with the data is relatively good for both
spectral function models, and suggests that at these kinematics
the description in terms of the smearing functions, with indications
of small but nonzero off-shell corrections, can provide a reliable
framework for describing electron scattering from $^3$He.
The agreement of the calculation with the data at kinematics
$Q_0^2 \approx 2$~GeV$^2$ comparable to some of the Day~{\it et al.}
spectra again suggests potential issues with these data.

\begin{figure}[t]
\includegraphics[width=16cm]{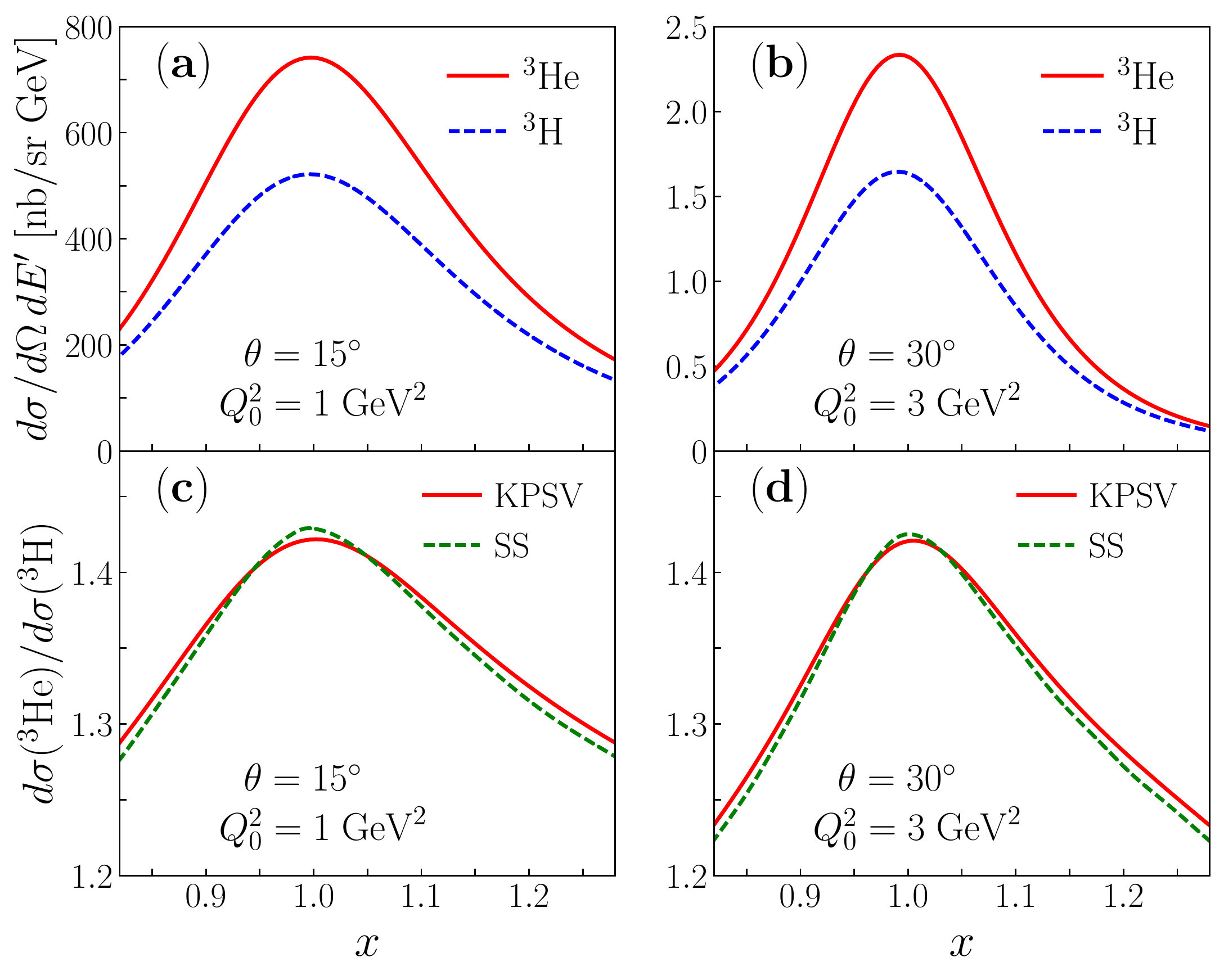}
\vspace*{-0.5cm}
\caption{QE cross section for $^3$He (red solid curves) and
	$^3$H (blue dashed curves) [in units of nb/sr$\cdot$GeV]
	at typical kinematics of the E12-11-112 experiment
	\cite{E12-11-112} with $E=4.3$~GeV,
	for (a) $\theta=15^\circ$ and $Q_0^2 = 1$~GeV$^2$
	and (b) $\theta=30^\circ$ and $Q_0^2 = 3$~GeV$^2$.
	The corresponding ratios of the $^3$He and $^3$H cross sections
	in (c) and (d) illustrate the effects of the different $A=3$
	spectral functions, from the KPSV (red solid curves) and
	SS (green dashed curves) models.}
\label{fig:QE_3He3H}
\end{figure}

In the near future the recently completed Jefferson Lab E12-11-112
experiment~\cite{E12-11-112} will provide additional information
on QE scattering in the $Q^2 \sim 1-3$~GeV$^2$ for both $^3$He
and $^3$H nuclei.
An estimate of the anticipated cross sections at the E12-11-112
kinematics is given in Fig.~\ref{fig:QE_3He3H}, for a beam energy
of $E=4.3$~GeV and scattering angles of $15^\circ$ and $30^\circ$.
Interestingly, the $^3$He cross section at the QE peak is
$\sim 30\% - 40\%$ larger than the corresponding $^3$H cross section,
which can be understood from the larger elastic contribution to the
proton structure function compared with the neutron,
	$F_2^{p \, (\rm el)} > F_2^{n \, (\rm el)}$,
which is doubly represented in $^3$He.
The wave function model dependence is again relatively weak,
as Fig.~\ref{fig:QE_3He3H} illustrates with the ratios of the
$^3$He to $^3$H cross sections.
Confronting these predictions with the E12-11-112 data will
provide important guidance for the identification of isospin
dependent effects in scattering from $A=3$ nuclei, and the
limitations of the impulse approximation and the WBA framework
for computing the smearing functions.

\subsection{Elastic form factors from QE data}

If the $^3$He and $^3$H smearing functions are sufficiently well
constrained at $y \approx 1$, the QE\ $^3$He and $^3$H data can
also be used to extract information about the nucleon's elastic
electromagnetic form factors.
In particular, from the ratio of $^3$He to $^3$H QE cross sections
measured in the E12-11-112 experiment~\cite{E12-11-112} at $x=1$,
and input on the proton's electromagnetic form factors and the
neutron's electric form factor, one can in principle extract the
free neutron's magnetic form factor, $G_{Mn}$.

A simple inversion of the QE formulas in the on-shell approximation
in Eqs.~(\ref{eq:F12QE}) allows the nucleon elastic form factors
$G_{EN}$, $G_{MN}$ to be determined from the smearing functions
$f_{ij}^N$ and the $^3$He to $^3$H structure functions.
To maximize the rates and simplify the analysis, one can take the
QE cross section in the vicinity of the QE peak, $x \approx 1$.
Taking the $F_2$ structure function as an example, the ratio of
the QE $^3$He to $^3$H functions can then be written
\begin{eqnarray}
R^{\rm (QE)}\
\equiv\ 
\frac{F_2^{^3{\rm He (QE)}}}{F_2^{^3{\rm H (QE)}}}
&=& \frac{2 + (f^n/f^p) R_{np}}{(f^n/f^p) + 2 R_{np}},
\label{eq:R3QE}
\end{eqnarray}
where $f^N \equiv f_{22}^N(x=1,\gamma)$ and
\begin{eqnarray}
R_{np} &=& \frac{G_{En}^2 + \tau G_{Mn}^2}{G_{Ep}^2 + \tau G_{Mp}^2}
\label{eq:Rnp-el}
\end{eqnarray}
is the ratio of the neutron to proton form factor combination
entering the $F_2$ structure function.
From Figs.~\ref{fig:fyp} and \ref{fig:fyn}, the ratio of the
neutron to proton smearing functions at $y=1$ is $\approx 0.87$,
almost independent of $\gamma$ for the range $\gamma=1-4$ considered
there, for both the KPSV and SS spectral function models.
The weak model dependence of the ratio is also illustrated in the
QE\ $^3$He to $^3$H cross section ratio in Fig.~\ref{fig:QE_3He3H},
which is $\approx 1.4$ at the QE peak.

Note that a slightly different combination of form factors would
be extracted from ratios of the QE\ $F_1^{A {\rm (QE)}}$ structure
functions, or from ratios of the actual cross sections, which are
combinations of $F_1^{A {\rm (QE)}}$ and $F_2^{A {\rm (QE)}}$.
In practice, this would be immaterial, as one could either extract
$F_1^{A {\rm (QE)}}$ and $F_2^{A {\rm (QE)}}$ from the cross section
by performing a Rosenbluth separation, or simply work in terms of a
different combination of the $G_{EN}$ and $G_{MN}$ form factors which
enters the cross section.

\begin{figure}[t]
\includegraphics[width=0.6\textwidth]{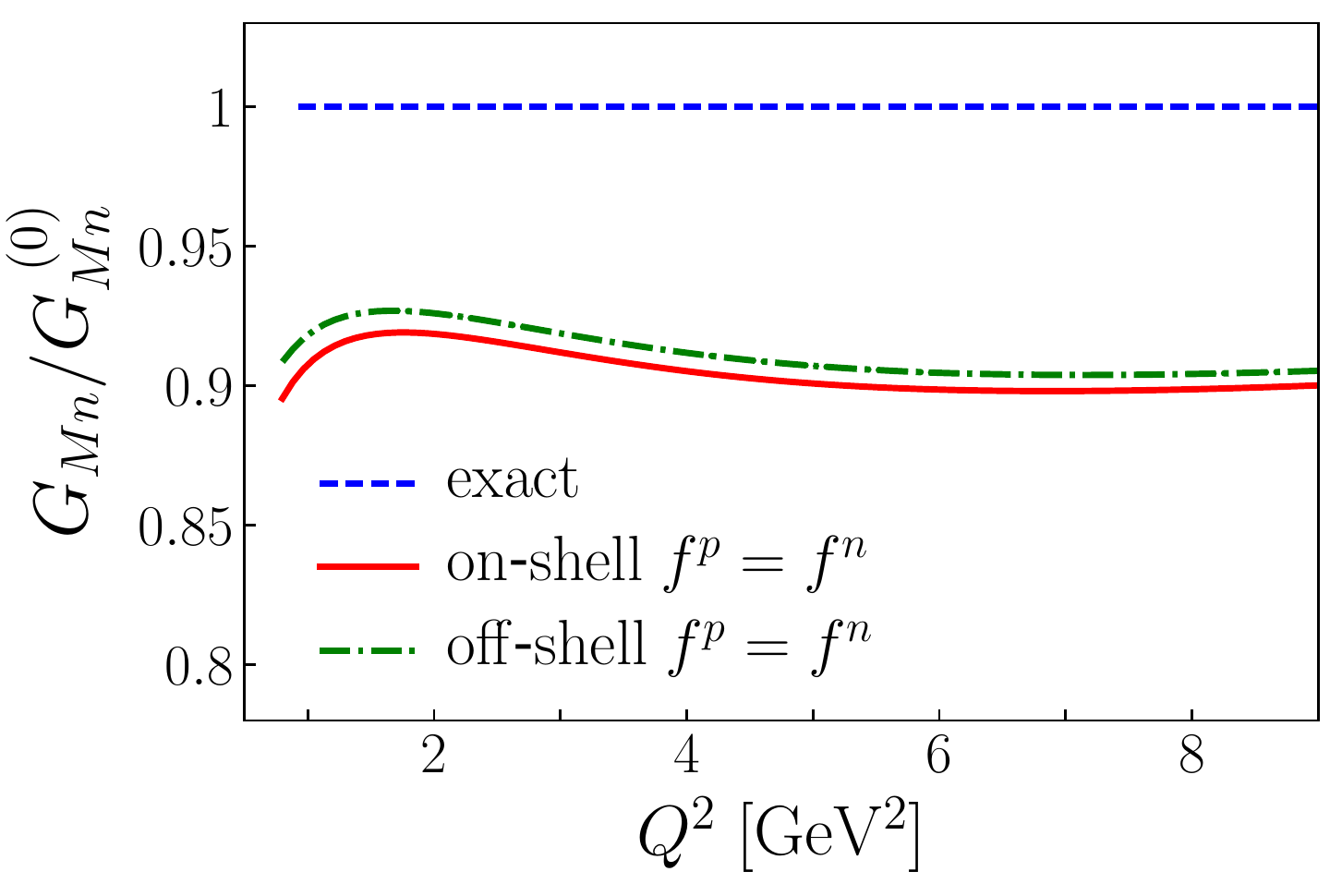}
\vspace*{-0.5cm}
\caption{Ratio of the neutron magnetic form factor $G_{Mn}$,
	extracted from Eqs.~(\ref{eq:Rnp-el}) and (\ref{eq:Rnp-R3QE}),
	to the input form factor $G_{Mn}^{(0)}$ taken from
	the Kelly parametrization~\cite{Kelly04}.
	The extracted $G_{Mn}$ is computed using the exact on-shell
	calculation (dashed blue curve), giving a ratio of unity,
	and with $G_{Mn}$ computed from Eq.~(\ref{eq:Rnp-R3QE})
	but with $f^p=f^n$ (solid red curve), or for $f^p=f^n$
	with the off-shell calculation using the ``cc2''
        prescription (dot-dashed green curve).}
\label{fig:A3_GMn}
\end{figure}

Inverting Eq.~(\ref{eq:R3QE}), one can write the form factor ratio
in Eq.~(\ref{eq:Rnp-el}) as
\begin{eqnarray}
R_{np} &=& \frac{(f^n/f^p) R^{\rm (QE)} - 2}
		{(f^n/f^p) - 2 R^{\rm (QE)}},
\label{eq:Rnp-R3QE}
\end{eqnarray}
with the QE ratio $R^{\rm (QE)}$ defined as in Eq.~(\ref{eq:R3QE}).
Measurement of $R^{\rm (QE)}$, together with a model for the
smearing function ratio $f^n/f^p$ and knowledge of $G_{Ep}$,
$G_{Mp}$ and $G_{En}$, can then be used to infer the magnetic
neutron form factor $G_{Mn}$.
In Fig.~\ref{fig:A3_GMn} we show the ratio of $G_{Mn}$ extracted
from Eq.~(\ref{eq:Rnp-R3QE}) to the input parametrization,
$G_{Mn}^{(0)}$.
The full calculation, illustrated here for the on-shell nucleon
structure function case with proton and neutron electromagnetic
form factors from Ref.~\cite{Kelly04}, of course gives a ratio of
unity, reflecting the self-consistency of the extraction method.
In contrast, if one were to use Eq.~(\ref{eq:Rnp-R3QE}) with
the assumption $f^p = f^n$, the extracted $G_{Mn}$ would be
$\approx 10\%$ lower over the range $Q^2 \approx 1 - 8$~GeV$^2$
than the true result.
Similar results are found for the off-shell calculation, as
illustrated in Fig.~\ref{fig:A3_GMn} for the ``cc2'' prescription.
Therefore, if one seeks experiments with precision for the extracted
$G_{Mn}$ to less than $\approx 10\%$ at these kinematics, using the
correct smearing function ratios would clearly be important in such
analyses.

\section{Deep-inelastic scattering from $^3$He and $^3$H}
\label{sec:dis}

The central motivation for the MARATHON experiment~\cite{MARATHON}
at Jefferson Lab is the measurement of the inclusive $^3$He to $^3$H
cross section ratio in the deep-inelastic scattering region, from
which one hopes to extract the ratio of the free neutron to proton
structure functions~\cite{Afnan00, Afnan03, Pace01, Sargsian02}.
At large values of $x$ ($x \gtrsim 0.6$) poor knowledge of the neutron
structure function has prevented a reliable determination of the $d/u$
quark PDF ratio in the proton from inclusive DIS data~\cite{MT96}.
Assuming that contributions from the scattering of longitudinal photons
are either sufficiently small or can be accurately estimated, the ratio
of the cross sections (\ref{eq:sigma}) for $^3$He and $^3$H can be used
to determine the $F_2^{^3\rm He}/F_2^{^3\rm H}$ structure function
ratio, from which $F_2^n/F_2^p$ can be extracted via~\cite{Afnan00,
Afnan03, Pace01, Sargsian02}
\begin{eqnarray}
\frac{F_2^n}{F_2^p}
&=& \frac{2 {\cal R} - F_2^{^3\rm He} / F_2^{^3\rm H}}
	 {2 F_2^{^3\rm He} / F_2^{^3\rm H} - {\cal R}},
\label{eq:fnfp}
\end{eqnarray}
where
\begin{eqnarray}
{\cal R} = \frac{R(^3\rm He)}{R(^3\rm H)}
\label{eq:super_R}
\end{eqnarray}
is the ``super-ratio'' of nuclear EMC ratios in $^3$He and in $^3$H,
\begin{subequations}
\label{eq:emc_ratios}
\begin{eqnarray}
R(^3\rm He) &=& \frac{F_2^{^3\rm He}}{2 F_2^p + F_2^n},	\\
\label{eq:emc_3he}
R(^3\rm H)  &=& \frac{F_2^{^3\rm H}}{F_2^p + 2 F_2^n}.
\label{eq:emc_3h}
\end{eqnarray}
\end{subequations}
Without a direct measurement of $F_2^n$, the EMC ratios $R(^3\rm He)$
and $R(^3\rm H)$ themselves cannot be uniquely determined.
However, irrespective of the magnitude of the nuclear corrections in
either $^3$He or $^3$H, if these effects are similar in the mirror
nuclei or can be reliably determined theoretically, then the uncertainty
introduced in the extraction of $F_2^n/F_2^p$ in Eq.~(\ref{eq:fnfp})
due to the super-ratio ${\cal R}$ can be minimized.

Several previous studies have estimated the super-ratio within
various nuclear models.
Among the standard approaches based on the impulse approximation,
Pace {\it et~al.}~\cite{Pace01} used a similar convolution framework
to that in Sec.~\ref{sec:theory}, together with smearing functions
computed in a correlated hyperspherical harmonics basis, including
Coulomb and three-body interactions.
Afnan {\it et~al.}~\cite{Afnan00, Afnan03} evaluated the super-ratio
in the convolution approximation using three-nucleon wave functions
obtained by solving the Faddeev equation, as well as using the
variational approach, while Sargsian {\it et al.}~\cite{Sargsian02}
employed a virtual-nucleon convolution model in addition to a model
based on light-front kinematics.
All these estimates found deviations of ${\cal R}$ of
$\lesssim 1\%-2\%$ from unity over the range accessible in the
MARATHON experiment.

Beyond the impulse approximation, Afnan {\it et~al.}~\cite{Afnan03}
considered the impact on the super-ratio of off-shell corrections
computed from a spectator quark model~\cite{MST94, Mulders92},
as well as from six-quark clusters, and a commonly used ansatz
based on nuclear density scaling~\cite{Frankfurt88}.
Sargsian {\it et al.}~\cite{Sargsian02} further considered a $Q^2$
rescaling model of the nuclear EMC effect~\cite{Close83, Close85},
and a color screening model in which off-shell effects were
represented in the form of short-range $NN$
correlations~\cite{Frankfurt88}.
To estimate the effect of possible isospin dependence of the
$NN$ correlation, the isosinglet and isotriplet combinations
were assumed to experience different amounts of suppression.
Of the scenarios considered, the isospin dependent effects
produced at most a $2\%-3\%$ deviation in the super-ratio for
$x \lesssim 0.8$.
Of course, any evidence of stronger isospin breaking corrections
could induce larger effects on the super-ratio.

In this section we explore whether the data on the ratio of $^3$He
to deuterium DIS cross sections from the Jefferson Lab E03-103
experiment~\cite{Seely09}, that were taken after the earlier
studies~\cite{Afnan00, Afnan03, Pace01, Sargsian02} were performed,
are able to provide any constraints on the possible isospin
dependence of the nuclear corrections and hence ${\cal R}$.
Moreover, as an alternative to the super-ratio method (\ref{eq:fnfp})
described above, we propose a more robust extraction procedure which,
although requiring additional experimental inputs, does not rely on
any assumptions about $\cal R$.
Before this, however, we first review some recent phenomenological
attempts to extract the off-shell corrections from global QCD analyses.

\subsection{Nucleon off-shell corrections}

Recently several global analyses of deuteron DIS and other high energy
scattering data~\cite{CJ15, AKP17} have obtained nucleon off-shell
contributions to the deuteron $F_2$ structure function in
Eq.~(\ref{eq:yconv2}) phenomenologically by fitting the isoscalar
off-shell function $\delta f^0$ directly,
\begin{eqnarray}
F_2^{d \rm (off)}(x)
&=& \int dy\, \widetilde{f}^{N/d}_{22}(y,\gamma)
    \left[ F_2^p\Big( \frac{x}{y} \Big)
	 + F_2^n\Big( \frac{x}{y} \Big)
    \right]
    \delta f^0\Big( \frac{x}{y} \Big),
\label{eq:F2doff}
\end{eqnarray}
where $\delta f^0$ is related to the proton and neutron
off-shell functions in Eq.~(\ref{eq:Fexp1}) by
\begin{eqnarray}
\delta f^0(x)
&=& \frac{F_2^p(x)\, \delta f^p(x) + F_2^n(x)\, \delta f^n(x)}
	 {F_2^p(x) + F_2^n(x)}.
\label{eq:delf0}
\end{eqnarray}
(For simplicity here we have suppressed the $Q^2$ dependence
in the structure functions.)
Despite some differences in the fitted shapes of the off-shell
functions in the two analyses~\cite{CJ15, AKP17}, the overall
magnitude of the off-shell effects was found to be relatively
small for the isoscalar combination $\delta f^0$.
Assuming isospin independence of the off-shell corrections,
Kulagin and Petti (KP)~\cite{KP06} also fitted data on ratios
of structure functions of heavy nuclei to deuterium, extracting
a universal function
  $\delta f^N \equiv \delta f^0 = \delta f^p = \delta f^n$
that agreed with the shape of that in Ref.~\cite{AKP17}, but
was somewhat larger than that from the CJ15 analysis~\cite{CJ15}.

\begin{figure}[t]
\includegraphics[width=0.7\textwidth]{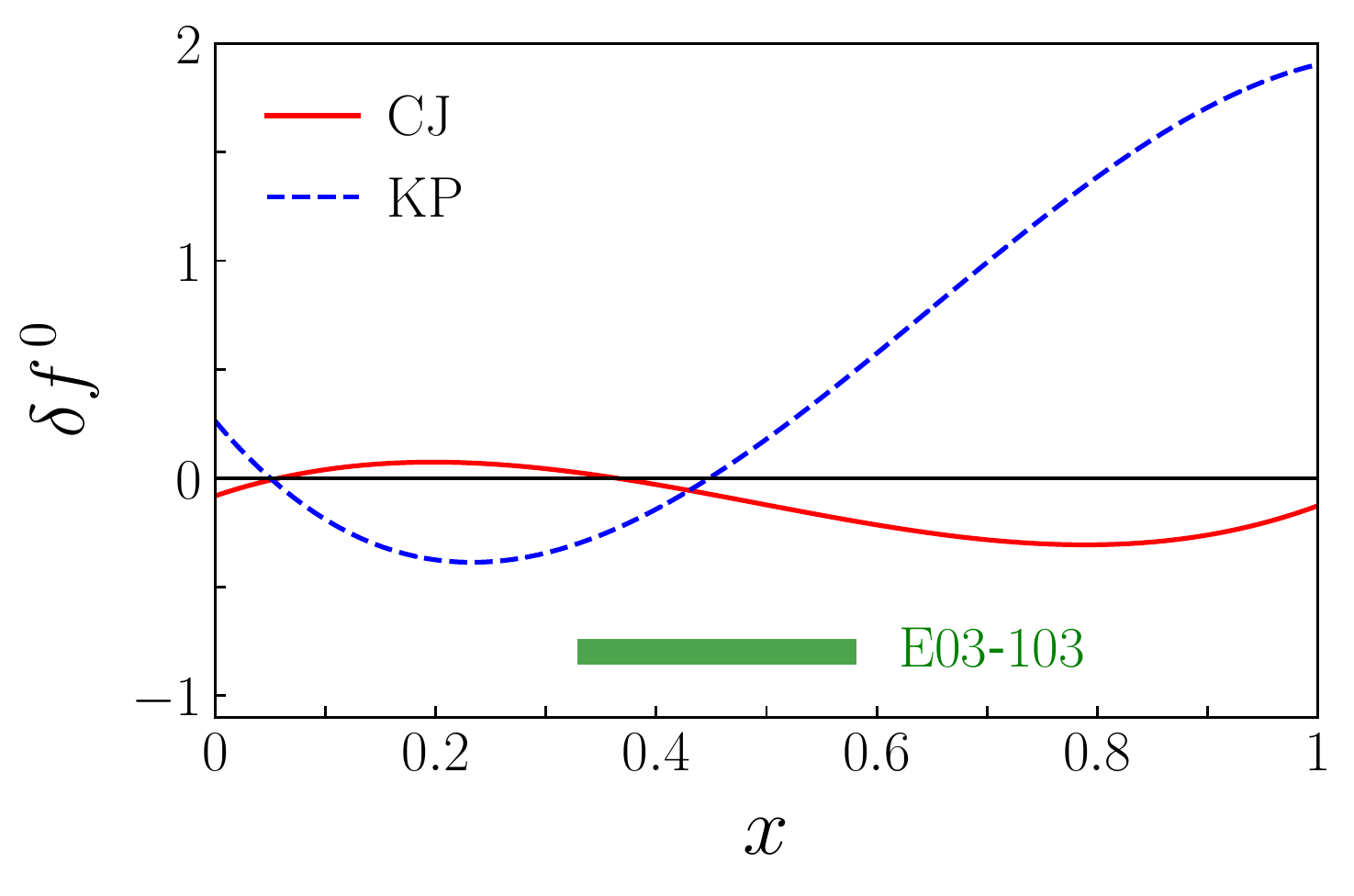}
\vspace*{-0.5cm}
\caption{Isoscalar off-shell function $\delta f^0$ from the
	CJ15~\cite{CJ15} global QCD analysis of proton and deuteron
	data (red solid curve) and the Kulagin-Petti~(KP)~\cite{KP06}
	fit to of nuclear structure function ratios, assuming
	$\delta f^p = \delta f^n$.  The kinematics of the Jefferson Lab
	E03-103 experiment~\cite{Seely09} are indicated by the green
	horizontal band.}
\label{fig:deltaf0}
\end{figure}

The isoscalar off-shell functions $\delta f^0$ from the CJ15 PDF
analysis~\cite{CJ15} and from the earlier KP nuclear structure
function fit~\cite{KP06} are shown in Fig.~\ref{fig:deltaf0}.
Both analyses used a parametrization based on a third order
polynomial of the form
\begin{eqnarray}
\delta f^0(x)
&=& C (x - x_0) (x - x_1) (1 + x_0 - x),
\label{eq:cj_form}
\end{eqnarray}
with parameters $x_0$ and $x_1$, and normalization $C$, which was
constructed to ensure at least one zero in the physical region of $x$.
The CJ15 analysis further imposed the normalization~\cite{CJ15}
\begin{eqnarray}
\int_0^1\!dx\, \delta f^0(x) \big[ q(x) - \bar{q}(x) \big] &=& 0
\label{eq:val_quark_constraint}
\end{eqnarray}
to ensure that the off-shell corrections do not modify the
valence quark number.
As Fig.~\ref{fig:deltaf0} illustrates, the CJ analysis found a
relatively small magnitude for $\delta f^0$, slightly positive
at low $x$ ($x \sim 0.2$) and negative at large $x$ ($x \gtrsim 0.4$).
The best fit corresponds to a deuteron wave function computed
from the AV18 $NN$ interaction~\cite{AV18}, although similar
quality fits were found using the CD-Bonn~\cite{CDBonn} and
WJC-2~\cite{WJC} wave functions, giving overall similar shapes
for $\delta f^0$.
In contrast, the off-shell function from the KP fit~\cite{KP06},
which uses the Paris $NN$ potential~\cite{Paris}, generally has
opposite sign compared with the CJ result in Fig.~\ref{fig:deltaf0},
and a somewhat larger magnitude which grows as $x \to 1$.
Interestingly, in the CJ analysis a similar shape to this was
found for the WJC-1~\cite{WJC} deuteron wave function, which,
however, gave a slightly larger overall $\chi^2$ value for the
global fit.

While the origin of the different behaviors for $\delta f^0$ found
in the two analyses is difficult to determine uniquely, one can
speculate that it may arise partly from the use of heavy nuclear
data in \cite{KP06}, which generally show a stronger nuclear
EMC effect than that in lighter nuclei.
To be conservative, in the present analysis we consider both
scenarios and investigate the consequences for the $A=3$
structure functions of both shapes for $\delta f^0$ shown
in Fig.~\ref{fig:deltaf0}.

\subsection{Isospin dependence of off-shell corrections}
\label{ssec:isospin-off}

Although deuterium data can only constrain the isoscalar combination
of PDFs and off-shell functions, data from the Jefferson Lab E03-103
experiment~\cite{Seely09} on the ratio of $^3$He to deuterium cross
sections could in principle allow the isospin dependence to be
disentangled.
In particular, because the $^3$He cross section is more sensitive to
proton structure, one can attempt to constrain the proton $\delta f^p$
correction from the $^3$He/$d$ ratio, and, using information from the
global analyses on $\delta f^0$~\cite{CJ15, KP06}, extract the neutron
off-shell correction from Eq.~(\ref{eq:delf0}),
\begin{eqnarray}
\delta f^n
&=& \frac{1}{F_2^n}
    \left[ (F_2^p + F_2^n) \delta f^0 - F_2^p\, \delta f^p
    \right]					\nonumber\\
&=& \delta f^0
 -  \frac{F_2^p}{F_2^n} (\delta f^p - \delta f^0).
\label{eq:deltafn}
\end{eqnarray}
In the remainder of this section we will analyze the $^3$He/$d$
data from Seely {\it et al.}~\cite{Seely09} within the theoretical
framework of Sec.~\ref{sec:theory}, and discuss the implications of
these data for the isospin dependence of the off-shell corrections.

\begin{figure}[t]
\includegraphics[width=\textwidth]{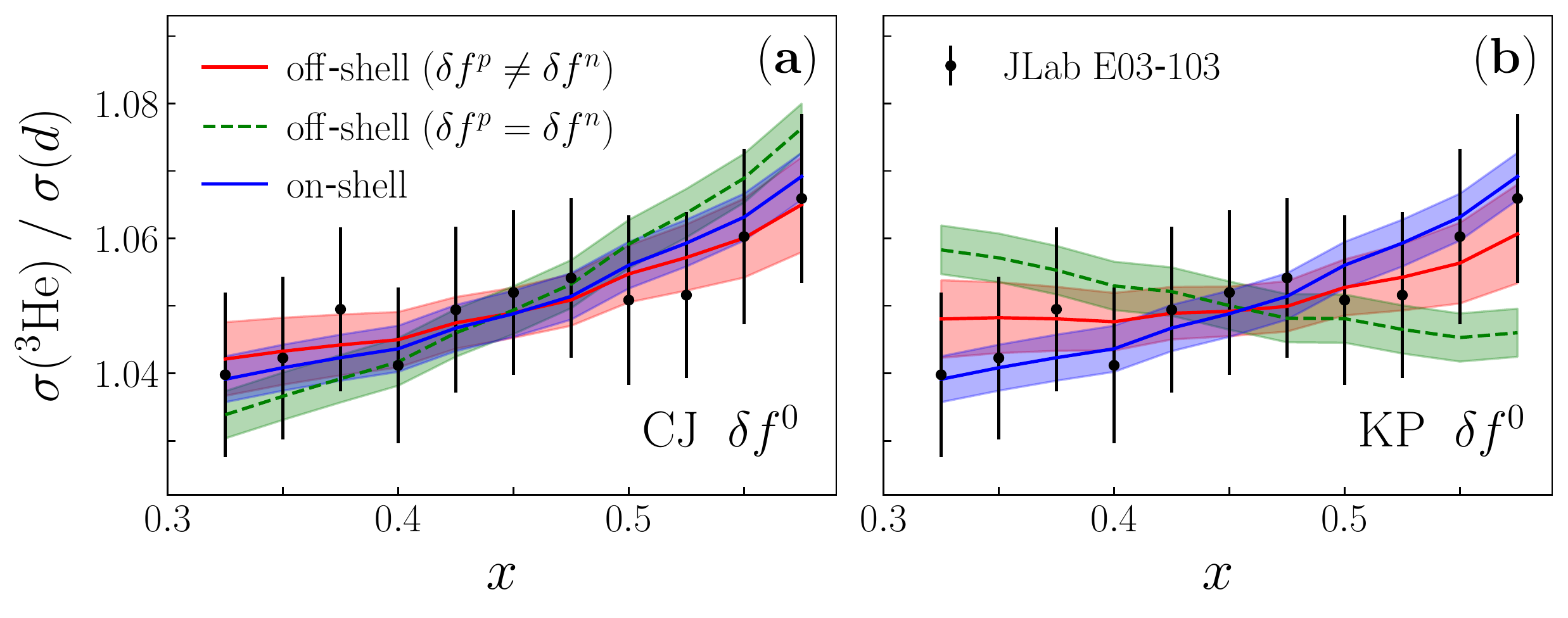}
\vspace*{-1cm}
\caption{Ratio of $^3$He to deuterium cross sections,
	$\sigma^{^3{\rm He}}/\sigma^d$, scaled per nucleon, from the
	Jefferson Lab E03-103 experiment~\cite{Seely09}	compared with
	the full Monte Carlo fit results (red solid curves and bands)
	using (a) the CJ~\cite{CJ15} and (b) the KP~\cite{KP06}
	isoscalar off-shell function $\delta f^0$, as well as with
	fits assuming isospin symmetric off-shell corrections,
	$\delta f^p = \delta f^n$ (green dashed curves and bands),
	and with the on-shell only fits (blue solid curves and bands).
	The experimental error bars include statistical and systematic
	uncertainties added in quadrature, with an overall 1.84\%
	normalization uncertainty not shown~\cite{Seely_data}.}
\label{fig:Seely}
\end{figure}

The $^3$He/$d$ data from the E03-103 experiment~\cite{Seely09}
were taken in Jefferson Lab Hall~C, using a 5.767~GeV beam of
electrons scattering mostly to an angle of $40^\circ$.
In the DIS region, $W^2 > 4$~GeV$^2$, the kinematics covered
the range $0.33 \lesssim x \lesssim 0.58$ and
$2.9 \lesssim Q^2 \lesssim 4.4$~GeV$^2$.
The measured ratio of the $^3$He to $d$ cross sections is shown in
Fig.~\ref{fig:Seely}, where the cross sections are scaled to those
per nucleon (total cross section ratio multiplied by a factor 2/3).
Note that the data here do not include any ``isoscalar correction'',
which can introduce unnecessary theoretical bias into the analysis.
The experimental error bars include statistical uncertainties and
point-to-point systematic uncertainties added in quadrature.
In addition, there is an overall 1.84\% fractional normalization
uncertainty that is not shown in Fig.~\ref{fig:Seely}.

For the analysis of the $^3$He/$d$ ratio, we fit the proton off-shell
function $\delta f^p$ using the same parametrization as for the
isoscalar off-shell function in Eq.~(\ref{eq:cj_form}).
Using the maximum likelihood method with Hessian error propagation,
we fit the $x$-intercept $x_0$ and the normalization parameter $C$,
and determine the position of the zero crossing at $x_1$ from the
off-shell normalization constraint (\ref{eq:val_quark_constraint}).
The results are found to be rather strongly dependent on the starting
parameters of the fit, indicating the presence of more than a single
$\chi^2$ minimum in parameter space.
To avoid this problem, we turn instead to a Monte Carlo analysis
method, using the nested sampling algorithm~\cite{Skilling:2004,
Mukherjee:2005wg, Shaw:2007jj} to map the likelihood function
into a Monte Carlo weighted parameter sample.
This method accounts for the possible presence of multiple minima,
and allows a rigorous determination of the fit uncertainties.
Similar methodology was recently used by the JAM Collaboration
to extract collinear PDFs~\cite{Sato:2016tuz, Ethier:2017zbq,
Barry:2018ort} and fragmentation functions~\cite{Sato:2016wqj},
as well as the transverse momentum dependent transversity
distribution~\cite{Lin:2017stx}.

\begin{figure}[t]
\includegraphics[width=0.9\textwidth]{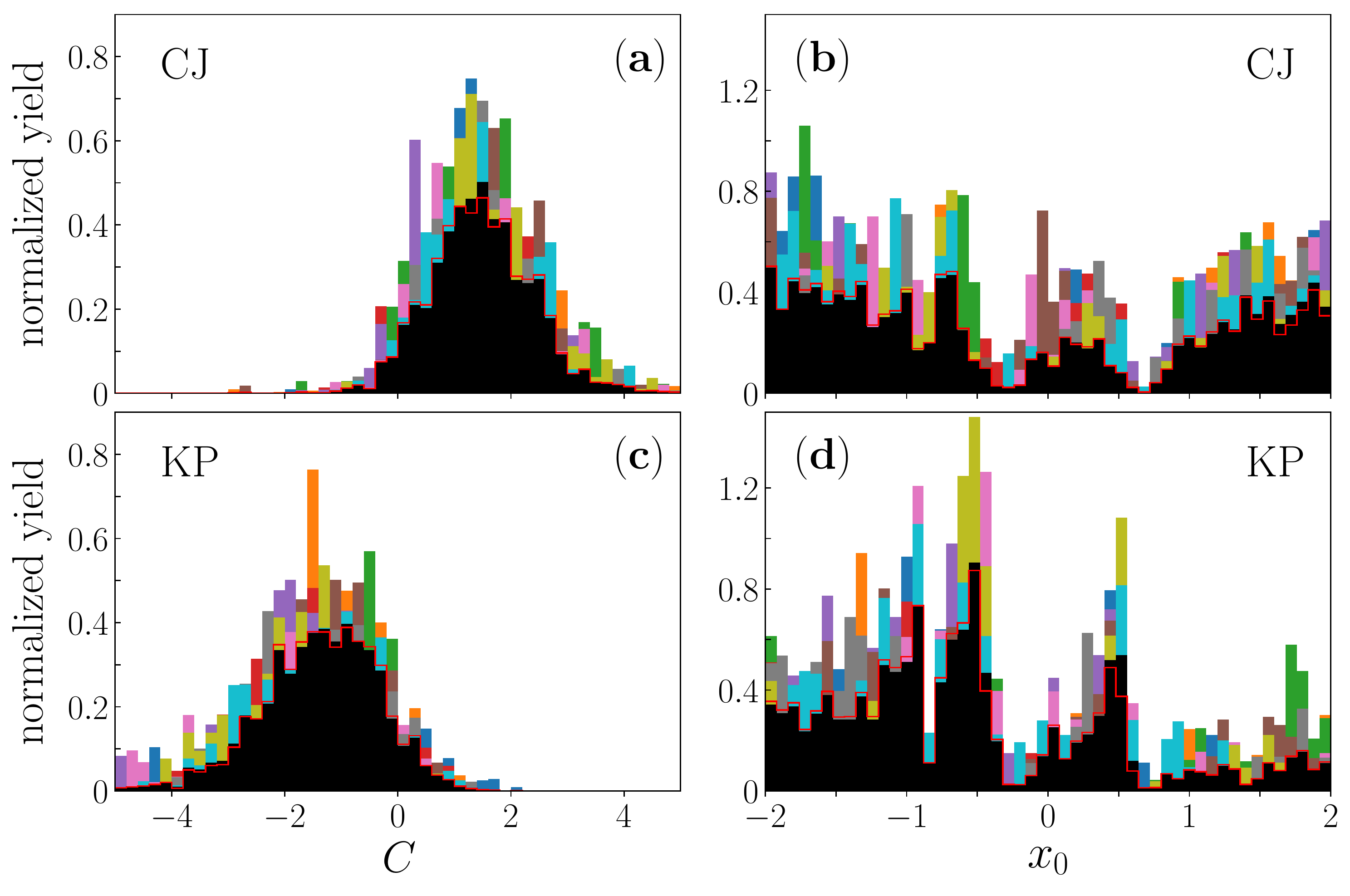}
\vspace*{-0.3cm}
\caption{Normalized yield of the Monte Carlo parameter distributions
	for the proton off-shell function $\delta f^p$, for the
	normalization~$C$ [(a) and (c)] and
	intercept $x_0$	[(b) and (d)],
	using the isoscalar off-shell function $\delta f^0$ from the
	CJ~\cite{CJ15} [(a) and (b)] and
	KP~\cite{KP06} [(c) and (d)] analyses.
	The colored histograms represent 10 statistically
	independent Monte Carlo analyses, while the black
	histograms are the combined result.}
\label{fig:params}
\end{figure}

\begin{figure}[t]
\includegraphics[width=\textwidth]{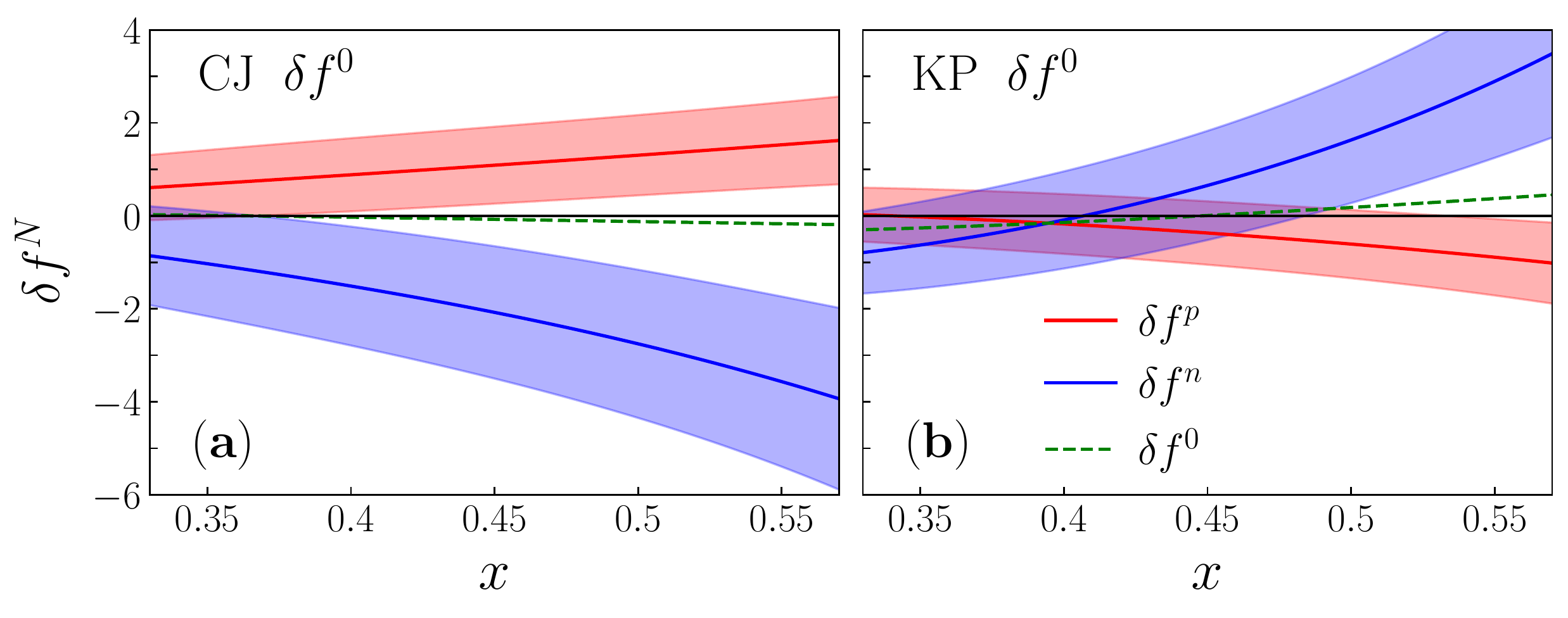}
\vspace*{-1cm}
\caption{Off-shell functions for the proton, $\delta f^p$
	(red solid curves and bands), and neutron, $\delta f^n$
	(blue solid curves and bands), from the fit to the E03-103
	data~\cite{Seely09}, for a given isoscalar off-shell
	function, $\delta f^0$ (green dashed), from
	(a) the	CJ~\cite{CJ15} global analysis and
	(b) the	Kulagin-Petti~\cite{KP06} nuclear ratios fit.
	The functions are shown only in the range of $x$
	constrained by the E03-103 data.}
\label{fig:deltaf}
\end{figure}

The results of the Monte Carlo fit to the $^3$He/$d$ data in
Fig.~\ref{fig:Seely} for the distribution of the fitted parameters
$C$ and $x_0$ are shown in Fig.~\ref{fig:params}.
For the CJ isoscalar function $\delta f^0$, the normalization
parameter $C$ is peaked for positive values, while the intercept
$x_0$ shows multiple solutions, both for $x_0 > 0$ and $x_0 < 0$.
This clearly illustrates the necessity for a Monte Carlo approach,
which can sample multiple solutions over a much larger range of
parameter space.
For the KP off-shell function, the distribution for the normalization
$C$ is generally confined to negative values, while the solutions for
$x_0$ are strongly peaked and appear to be somewhat anticorrelated
with the values found for the CJ result.

From these fitted parameters, the resulting off-shell functions
$\delta f^p$ and $\delta f^n$ are computed in Fig.~\ref{fig:deltaf},
for both the CJ and KP off-shell isoscalar distributions $\delta f^0$.
In the kinematic region constrained by the \mbox{E03-103} data,
$0.3 \lesssim x \lesssim 0.6$, the proton off-shell function
$\delta f^p$ is found to be positive and significantly larger than
the isoscalar function $\delta f^0$ for both CJ and KP fits.
Consequently, from Eq.~(\ref{eq:deltafn}) the neutron off-shell
function $\delta f^n$ becomes negative [see the second term in
(\ref{eq:deltafn})], with its magnitude enhanced by the factor
$F_2^p/F_2^n > 1$.
For the CJ fit the proton and neutron off-shell functions remain
positive and negative, respectively, over the fitted range, while
for the KP off-shell fit there is a sign change at $x \approx 0.4$.

Although the absolute values of the proton and neutron off-shell
functions in Fig.~\ref{fig:deltaf} are large relative to the
isoscalar functions, the respective contributions to the nucleon
structure functions are weighted by the nucleon virtuality
$v(p^2) \ll 1$.
For $^3$He, the average proton and neutron virtualities
(for the KPSV spectral function) are found to be $\approx -7\%$
and $\approx -9\%$, respectively.
At the lower end of the $x$ range covered by the experiment,
the relative correction to the nucleon structure functions are
$\lesssim 10\%$, and rise to $\sim 30\%$ for the neutron at the
higher $x$ values.
Off-shell corrections that are very large $(\gtrsim 30\%-50\%)$
are likely to invalidate the lowest order expansion in $v$ assumed
in Eq.~(\ref{eq:Fexp1}), or may suggest issues with the systematic
uncertainties assigned to the $^3$He/$d$ data~\cite{Seely09}.

\begin{figure}[t]
\includegraphics[width=0.85\textwidth]{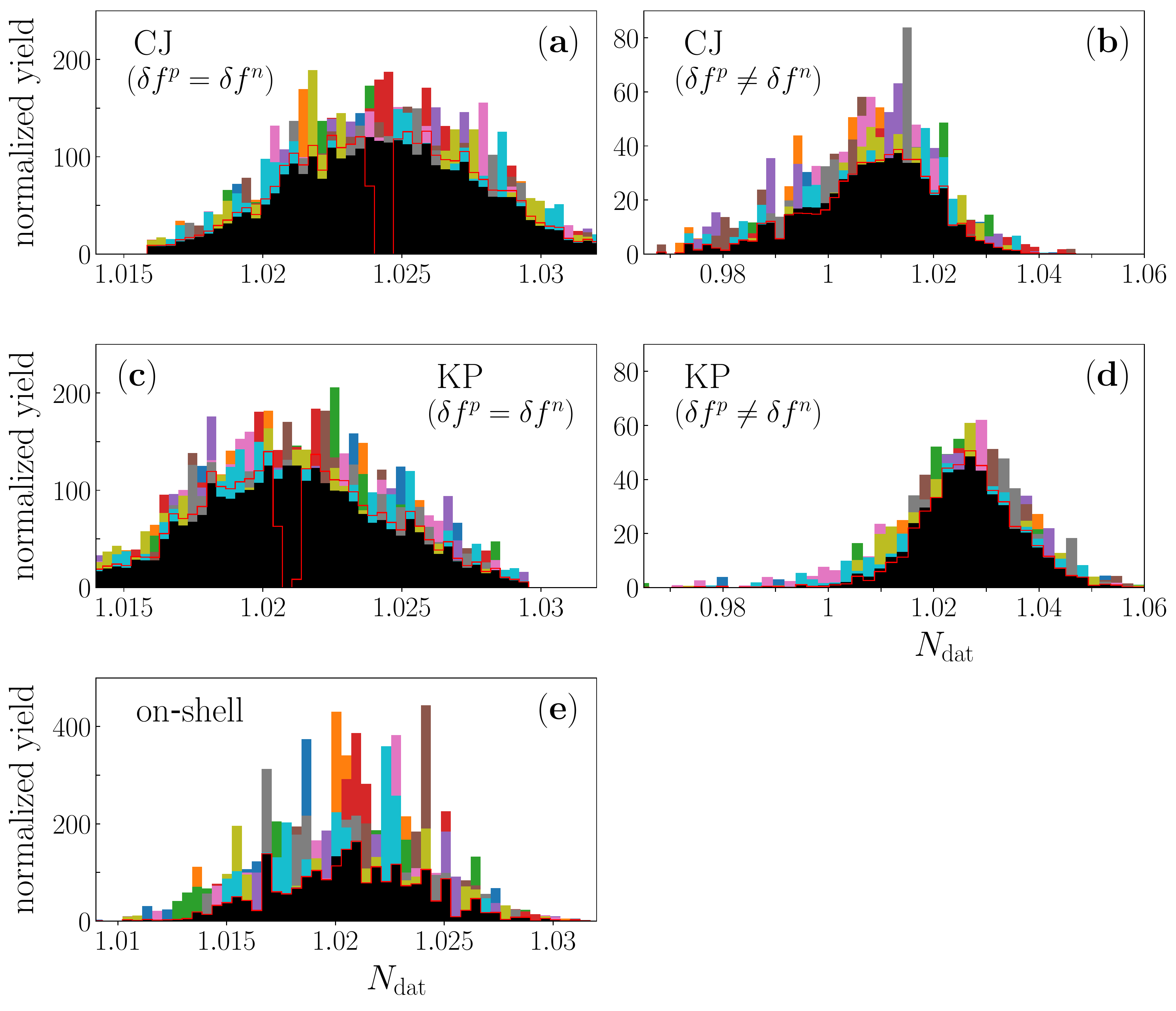}
\vspace*{-0.3cm}
\caption{Normalized yield of Monte Carlo distributions for the
	data normalization factors $N_{\rm dat}$ for the
	CJ~\cite{CJ15} [(a) and (b)] and
	KP~\cite{KP06} [(c) and (d)] isoscalar
	functions $\delta f^0$, assuming isospin symmetry
	($\delta f^p = \delta f^n$) [(a) and (c)]
	and the isospin dependent analysis
	($\delta f^p \not= \delta f^n$) [(b) and (d)],
	along with the on-shell only fit [(e)].
	The colored histograms represent 10 statistically
	independent Monte Carlo analyses, while the black
	histograms are for the combined result.
	The pile-up in some of the fits occurs at the upper
	boundary of the allowed 1.84\% normalization
	uncertainty in the E03-103 experiment~\cite{Seely09}.}
\label{fig:norm}
\end{figure}

Taking into account the overall normalization uncertainty of the
$^3$He/$d$ ratio data, the Monte Carlo distribution of the fitted
data normalization parameter, $N_{\rm dat}$, is shown in
Fig.~\ref{fig:norm}, with values restricted to lie within the
1.84\% quoted for the E03-103 experiment~\cite{Seely09}.
For the fit using the CJ isoscalar function, the distribution
is relatively broad, with a peak at around 1\% and an average of
$N_{\rm dat} = 1.006 \pm 0.009$.  This gives a very good overall
fit to the E03-103 data, as evident from Fig.~\ref{fig:Seely}.
For the KP off-shell function, the normalization parameter
distribution is more concentrated at the upper limit, with an
average $N_{\rm dat} = 1.012 \pm 0.005$.  The resulting fit to
the $^3$He/$d$ data is not quite as good at the lower $x$ values,
but still consistent with the data within $1\sigma$.

Note that the full Monte Carlo fit clearly disfavors zero off-shell
corrections, $C=0$, especially for the CJ isoscalar function, since it
is easier for the fit to vary one of the free parameters than to keep
the same shape and compensate by a normalization shift in the data.
Nevertheless, if the off-shell corrections are switched off
``by hand'', one can still obtain a good fit to the $^3$He/$d$
data with just the on-shell contributions, as illustrated in
Fig.~\ref{fig:Seely}, with an average data normalization shift
$N_{\rm dat} = 1.016 \pm 0.002$, consistent with the maximum
1.84\% allowed [see Fig.~\ref{fig:norm}].
The $\chi^2/{\rm dof}$ value for the on-shell fit is slightly larger
than that for the off-shell fit, but is still $< 1$ and within
1$\sigma$ from the best fit, even though the tendency is towards
a shape with a slightly different slope than the data prefer.

In an earlier analysis of the E03-103 $^3$He/$d$ ratio,
Kulagin and Petti showed~\cite{KP10} that with the KP off-shell
correction, and assuming $\delta f^p = \delta f^n$, one could fit
the Seely {\it et al.} data with a 3\% normalization shift, and
be consistent with extractions of $F_2^n/F_2^p$ from NMC data.
This value lies outside of the $1\sigma$ range for $N_{\rm dat}$
quoted by the experiment.
Using our Monte Carlo methodology, we also attempt to fit the
E03-103 data using the isospin symmetric KP off-shell function.
Constraining the normalization $N_{\rm dat}$ to be within the
quoted experimental uncertainty range, the fit shown in
Fig.~\ref{fig:norm} prefers the maximum upward shift of the
data, with an average value $N_{\rm dat} = 1.016 \pm 0.001$.
The resulting $\sigma^{^3{\rm He}}/\sigma^d$ ratio does not
give as good a description of the data in Fig.~\ref{fig:Seely},
overestimating the ratio at lower $x$ and underestimating it
at higher $x$.
If one uses instead the CJ isoscalar off-shell correction,
assuming isospin symmetry, the fitted data normalization is
also near the maximum allowed, $N_{\rm dat} = 1.017 \pm 0.001$.
The resulting fit to the $^3$He/$d$ data in Fig.~\ref{fig:Seely}
shows good agreement at lower $x$, but overestimates the
data at the higher $x$ values.

The inescapable conclusion is that, taking the E03-103
$^3$He/$d$ data~\cite{Seely09} with the quoted uncertainties
at face value, the fits clearly disfavor isospin symmetric
off-shell corrections, and slightly favor isospin dependent
off-shell effects over no off-shell corrections.
In the next section we examine the consequences of this for
the MARATHON experiment and the extraction of the neutron to
proton structure function ratio.

\subsection{Implications for $A=3$ structure functions}
\label{ssec:implications}

Having obtained the constraints on the nucleon off-shell functions
from the E01-103\ $^3$He/$d$ data~\cite{Seely09} and the previously
determined isoscalar off-shell function, we next discuss the
implications of these results for the structure functions of
$A=3$ nuclei.
In particular, the MARATHON experiment~\cite{MARATHON} at Jefferson Lab
will make high-precision measurements of the inclusive cross section
ratios for $^3$He to $^3$H, as well as $^3$He/$d$ and $^3$H/$d$,
which are expected to yield information on the ratio of the free
neutron to proton structure functions.
If one uses the super-ratio method in Eq.~(\ref{eq:fnfp}), the effect
of the off-shell corrections extracted in Sec.~\ref{ssec:isospin-off}
on the ${\cal R}$ ratio will therefore be of direct relevance for the
$n/p$ determination.

\begin{figure}[t]
\includegraphics[width=\textwidth]{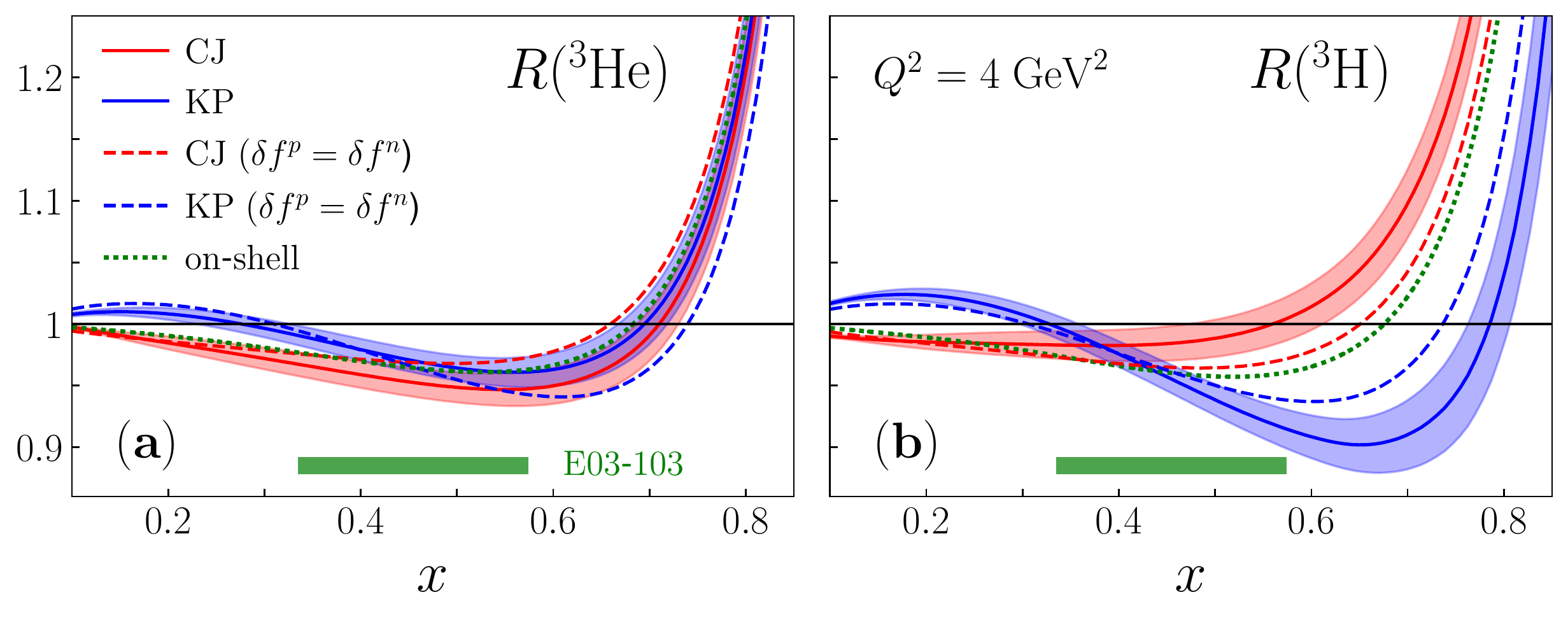}
\vspace*{-1cm}
\caption{Ratios of nuclear to nucleon structure functions
	(a) $R(^3{\rm He}) = F_2^{^3{\rm He}}/(2 F_2^p + F_2^n)$ and
	(b) $R(^3{\rm H})  = F_2^{^3{\rm H} }/(F_2^p + 2 F_2^n)$,
	for the off-shell Monte Carlo fits using the
	CJ~\cite{CJ15} (solid red curves and bands) and
	Kulagin-Petti~\cite{KP06} (solid blue curves and bands)
	isoscalar off-shell function $\delta f^0$, compared with
	the ratios assuming isospin symmetric off-shell functions
	from CJ (dashed red curves) and KP (dashed blue curves),
	and the on-shell only fit (dotted green curves).
	The range of $x$ constrained by the Jefferson Lab E03-103
	experiment~\cite{Seely09} is indicated by the green
	horizontal band, and a scale of $Q^2=4$~GeV$^2$, which
	is close to the average for the E03-103 data, was used
	for all structure functions.}
\label{fig:EMCratios}
\end{figure}

For the on-shell only calculation, Fig.~\ref{fig:EMCratios} shows
rather similar $^3$He and $^3$H EMC ratios, with both $R(^3\rm He)$
and $R(^3\rm H)$ having minima at $x \approx 0.5-0.6$, at which
they dip $\approx 4\%-5\%$ below unity, before rising rapidly at
$x \gtrsim 0.7$ through Fermi motion.
Because of the greater sensitivity of the $^3$He and $^3$H ratios
to any isospin dependence of off-shell effects, including the
off-shell corrections from Fig.~\ref{fig:deltaf} gives rise to some
quite interesting features.
Since the $^3$He ratio is more sensitive to proton structure than to
the neutron, for the case of the CJ isoscalar off-shell correction
the fitted positive proton off-shell function $\delta f^p$ induces a
slightly stronger EMC effect, with the dip in $R(^3\rm He)$ increasing
to $\approx 5\%$.
In contrast, since the neutron plays a greater role in the $^3$H EMC
ratio, the fitted negative neutron off-shell correction $\delta f^n$
reduces the dip in $R(^3\rm H)$ to $\lesssim 2\%$ for $x \lesssim 0.5$,
with an earlier onset of the Fermi motion rise above unity.

For the KP isoscalar off-shell function $\delta f^0$, which gives
similarly small fitted proton and neutron off-shell corrections at
$0.3 \lesssim x \lesssim 0.4$, but increasing magnitudes for the
(positive) neutron $\delta f^n$ and (negative) proton $\delta f^p$
at larger $x$, the effect on the $^3$He and $^3$H EMC ratios is
more dramatic.
In particular, the positive neutron off-shell function enhances
the magnitude of the dip in the $R(^3\rm H)$ ratio to almost 10\%
at $x \approx 0.65$, in marked contrast to the prediction with
the CJ $\delta f^0$.
The impact on the $R(^3\rm He)$ ratio is much less at large $x$,
with little deviation of the KP off-shell result from the on-shell
fit at $x \gtrsim 0.4$.
At smaller $x$ values, $x \lesssim 0.3$, the KP off-shell corrections
yield an enhancement of $\approx 2\%-3\%$ above unity in both the
$^3$He and $^3$H ratios, which is directly related to the dip
in the KP $\delta f^0$ function at $x \approx 0.2$ seen in
Fig.~\ref{fig:deltaf0}.
On the other hand, there is currently no compelling evidence for
such an enhancement from deuterium data \cite{CJ15}, and the effect
in the KP $\delta f^0$ may be due to the use of data on heavy nuclei
in the KP analysis \cite{KP06}, which do display some enhancement
of $F_2^A/F_2^d$ at $x \sim 0.1-0.2$.

Note also that the off-shell corrections are constrained by
the E03-103 data~\cite{Seely09} only in the range between
$x \approx 0.3$ and 0.6, and outside this range, where the
low-$x$ enhancement for the KP case and the growing differences
between the $R(^3{\rm H})$ ratios at large $x$ are apparent,
these are not directly constrained by data.
Measurement of the $^3$H structure function in the MARATHON
experiment \cite{MARATHON}, covering a wide range of $x$ values,
$0.2 \lesssim x \lesssim 0.8$, will provide an unprecedented
opportunity to examine the role of nucleon off-shell effects in
the $A=3$ system, as well as their possible isospin dependence.
In fact, as Fig.~\ref{fig:EMCratios} illustrates, the $R(^3{\rm He})$
and $R(^3{\rm H})$ ratios show sensitivity to the off-shell
corrections even if these are assumed to be isospin symmetric,
	$\delta f^0 = \delta f^p = \delta f^n$.
In the case of $\delta f^0$ determined from the CJ analysis \cite{CJ15},
the off-shell corrections move both the $^3$He and $^3$H ratios
upward relative to the on-shell calculation, resulting in slightly
weaker EMC effects for both nuclei.
For $\delta f^0$ taken from the KP analysis \cite{KP06}, the effect
is a downward shift, making the EMC effects in $^3$He and $^3$H
slightly larger.
Consequently, the relative shifts in $R(^3{\rm He})$ and $R(^3{\rm H})$
in both models are similar.

\begin{figure}[t]
\includegraphics[width=\textwidth]{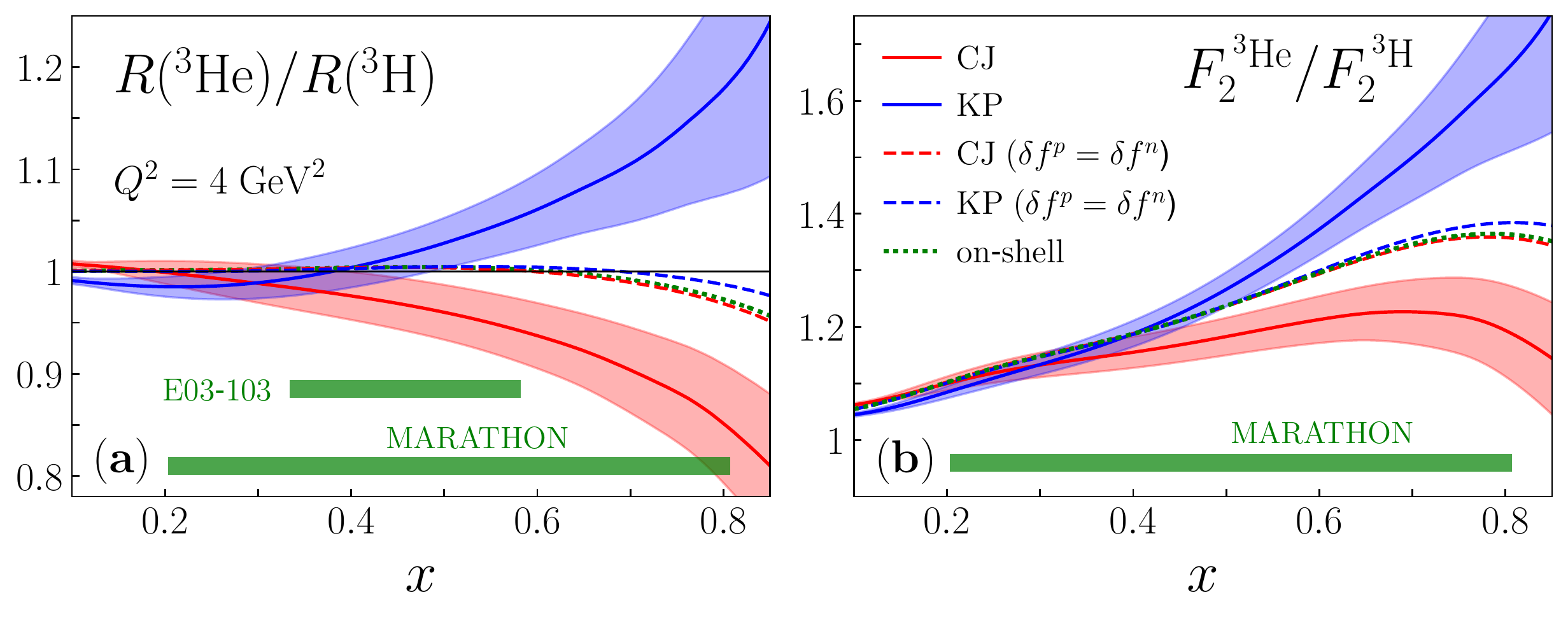}
\vspace*{-1cm}
\caption{(a) Super-ratio $R(^3{\rm He})/R(^3{\rm H})$
	of the EMC ratios in $^3$He and $^3$H and
	 (b) the ratio $F_2^{^3{\rm He}}/F_2^{^3{\rm H}}$
	of the $^3$He and $^3$H structure functions.
	The off-shell Monte Carlo fits using the CJ~\cite{CJ15}
	(solid red curves and bands) and Kulagin-Petti~\cite{KP06}
	(solid blue curves and bands) isoscalar off-shell functions
	$\delta f^0$ are compared with the results assuming isospin
	symmetric off-shell functions from CJ (dashed red curves)
	and KP (dashed blue curves), and the on-shell only fit
	(dotted green curves).
	The range of $x$ constrained by the Jefferson Lab E03-103
	experiment~\cite{Seely09} and the extended range expected
	by the MARATHON experiment~\cite{MARATHON} are indicated
	by the green horizontal bands.}
\label{fig:Super-ratios}
\end{figure}

This can be more directly seen in the super-ratio,
	${\cal R} = R(^3\rm He)/R(^3\rm H)$,
of the $^3$He and $^3$H EMC ratios, in Fig.~\ref{fig:Super-ratios}.
For both the on-shell only and isospin symmetric off-shell fits,
the super-ratio is within $\approx 1\%$ of unity for $x \lesssim 0.7$,
with the deviations increasing slightly at larger $x$ values.
(Recall, however, from Fig.~\ref{fig:Seely} that the fits to the
E03-103 data~\cite{Seely09} with the isospin symmetric off-shell
corrections give the worst agreement, especially for the KP
isoscalar correction.)
For the isospin asymmetric off-shell functions, the deviations
from unity are at the few-percent level up to $x \approx 0.4$, but
become significantly larger at higher $x$, reaching $\approx 15\%$
above unity at $x = 0.8$ for the CJ fit, and a similar amount
below unity for the KP result, albeit with large uncertainties.
This translates to a ratio of $^3$He to $^3$H structure functions,
which will be extracted from the MARATHON experiment, that deviates
from the on-shell result by up to $\approx -10\%$ for the CJ result
and $\approx +20\%$ for the KP fit at $x = 0.8$, with $\approx 50\%$
statistical uncertainties on these values.

Of course, as discussed above, the results on the off-shell
corrections are constrained by the E03-103 data only for
$x \lesssim 0.6$, above which their extrapolation cannot be
considered very reliable.
In the region specificically covered by the E03-103 experiment
\cite{Seely09}, the off-shell effects scale up to $\approx 5\%$,
although in opposite directions for the CJ and KP isoscalar
off-shell corrections.
Measurement of the $^3$He/$^3$H ratio in MARATHON would therefore
be vital for discriminating between these scenarios.

On the other hand, without additional assumptions it may be
difficult to attribute the differences such as those in
Fig.~\ref{fig:Super-ratios} entirely to different proton
and neutron off-shell corrections, or to a different behavior
of the free neutron structure function at large $x$.
In the following we discuss an alternative analysis scenario,
in which the MARATHON data on the nuclear structure function
ratios can be used as critical input for a simultaneous
determination of both the neutron to proton ratio {\it and}
the isospin dependence of the nucleon off-shell corrections.

\subsection{Extracting neutron structure from MARATHON}

While our Monte Carlo analysis suggests that the possibility
of strong isospin dependence of the nucleon off-shell effects at
high~$x$ cannot be ruled out on the basis of the \mbox{E03-103}
data~\cite{Seely09}, it is necessary to examine the caveats and
assumptions that underlie these findings.
Firstly, our extraction of the proton and neutron off-shell functions
$\delta f^p$ and $\delta f^n$ assumes the isoscalar nucleon off-shell
correction $\delta f^0$ to be reliably determined from previous analyses
of the proton and deuteron data (or, in the case of KP, also of heavy
nuclear structure function ratios).  However, as is obvious from the
sizeable differences between the CJ and KP $\delta f^0$ functions in
Fig.~\ref{fig:deltaf0} and in their consequences for the super-ratio
in Fig.~\ref{fig:Super-ratios}, the magnitude of the isoscalar off-shell
correction, and even its sign as a function of $x$, is controversial.
Futhermore, in our analysis we have used the same set of input nucleon
PDFs \cite{CJ15} with both the CJ and KP isoscalar off-shell functions.
While this is consistent for the CJ $\delta f^0$, for the analysis with
the KP off-shell function one should in principle use the PDF set that
was used in the extraction of $\delta f^0$ in Ref.~\cite{KP06}.
The KP analysis \cite{KP06} assumed, however, that
$\delta f^p = \delta f^n$ in the fits to structure function ratios
for asymmetric nuclei, so using the KP $\delta f^0$ to determine the
isospin dependence of $\delta f^N$ in our analysis is not entirely
consistent.

A more reliable determination of the proton and neutron off-shell
corrections would involve a {\it simultaneous} analysis of proton,
deuteron and $A=3$ nuclear data.  This would remove many systematic
effects arising from different theoretical assumptions and inputs
utilized in the different analyses.  Whatever tensions then remain
between data sets in the combined fit would be treated consistently
within the same analysis.
In principle, while a global QCD analysis is the most natural
framework in which to perform the simultaneous fit to the nucleon
PDFs and nuclear off-shell functions, one could also imagine a more
restricted fit at the structure function level.

In particular, with sufficient experimental information on the
structure functions of $^3$He, $^3$H and deuterium, one can in
practice disentangle the nuclear effects from the on-shell nucleon
structure functions.
Within the convolution framework of Sec.~\ref{ssec:wba}, the nuclear
structure functions are expressed as sums of on-shell and off-shell
contributions, as in Eq.~(\ref{eq:dis_onoff}),
\begin{eqnarray}
F_2^A(x,Q^2)
&=& F_2^{A\, \rm (on)}(x,Q^2)  +  F_2^{A\, \rm (off)}(x,Q^2),
\end{eqnarray}
for $A=d$, $^3$He and $^3$H.
The on-shell term depends on the free proton and neutron structure
functions, $F_2^p$ and $F_2^n$, and the nuclear smearing functions,
$f_{ij}^{N/A}$, in Eq.~(\ref{eq:fij}).
The latter are reasonably well determined away from the tails of
the distributions at large $y$, which become important only at
$x \sim 1$.
The off-shell term depends on $F_2^p$, $F_2^n$, $\delta f^p$,
$\delta f^n$ and the off-shell smearing functions
$\widetilde{f}_{ij}^{\, N/A}$ in Eq.~(\ref{eq:fij_off}), which are
computed in terms of the {\it same} set of nuclear wave functions
as the on-shell smearing functions $f_{ij}^{N/A}$.

Since the proton $F_2^p$ structure function is well known, the three
unknowns --- the $F_2^n/F_2^p$ ratio, and the two off-shell corrections,
$\delta f^p$ and $\delta f^n$ --- can then be determined from three
independent observables, such as the ratios $^3$He/$d$ and $^3$H/$d$
(or $^3$He/$^3$H) and $F_2^d/F_2^p$.
The ratios involving $^3$He and $^3$H are the main focus of the
MARATHON experiment; however, the experiment will also measure the
deuteron/proton structure function ratio over a more restricted
range of kinematics, from $x=0.18$ to $0.38$ (for $Q^2$ between
2.5 and 5.3~GeV$^2$), which will be used to benchmark against the
large body of high precision data on inclusive $F_2^d$ and $F_2^p$
structure functions that has been accumulated over the past few
decades~\cite{Shujie18}.

While in global QCD analyses one typically parametrizes individual
PDFs from which all observables are then constructed, for an analysis
at the structure function level the $x$ and $Q^2$ dependence of
the structure functions could be parametrized by a form such
as~\cite{Duke:1983gd}
\begin{align}
F_2(x,Q^2)
&= a_0(Q^2) x^{a_1(Q^2)} (1-x)^{a_2(Q^2)}
	    (1 + a_3(Q^2) \sqrt{x} + a_4(Q^2) x + \cdots),
\end{align}
with $Q^2$ dependent shape parameters
\begin{align}
a_i(Q^2) &= a_i^{(0)} + a_i^{(1)} s(Q^2),
\hspace*{1cm}
s(Q^2)\ = \log \left(\frac{\log(Q^2/\Lambda_{\rm QCD}^2)}
			  {\log(\mu^2/\Lambda_{\rm QCD}^2)}
	       \right),
\end{align}
for $i=1-4$, where $\Lambda_{\rm QCD}$ is the QCD scale parameter,
and $\mu^2$ is a scale of order ${\cal O}({\rm 1~GeV}^2)$ fitted
to the data.
For the proton and neutron off-shell functions a polynomial
of degree~3, as in Eq.~(\ref{eq:cj_form}), would be expected
to be sufficient.
(For simplicity one can assume that $\delta f^N$ is independent of
$Q^2$, so that the scale dependence of the off-shell contributions
to the structure functions is the same as the on-shell.)
The extraction of the three unknown functions would then involve
fitting $\sim 30$ parameters, which can be constrained within a
Bayesian likelihood analysis.
In this approach, recently employed by the JAM collaboration in
their Monte Carlo analyses of PDFs and fragmentation functions
\cite{Sato:2016tuz, Ethier:2017zbq, Barry:2018ort, Sato:2016wqj,
Lin:2017stx}, the multivariate probability density for a set of
fit parameters $\bm{a} = \{ a_i \}$ conditioned by the data
is given by
$p(\bm{a}|{\rm data})
  \propto {\cal L}({\rm data}|\bm{a})\, \pi(\bm{a})$,
where the likelihood ${\cal L}$ is a Gaussian function of the
$\chi^2$,
\begin{align}
{\cal L}({\rm data}|\bm{a})
= \exp \left[ -\frac{1}{2}\chi^2(\bm{a},{\rm data}) \right],
\end{align}
and $\pi(\bm{a})$ is the distribution of priors.
The $\chi^2$ function takes into account experimental statistical,
systematic (uncorrelated and correlated), and normalization
uncertainties for each data set used in the fit \cite{Sato:2016tuz}.
The expectation values and $1\sigma$ uncertainties for the fitted
quantities can then be computed by Monte Carlo sampling of the
probability density \cite{Sato:2016wqj}.

The remaining approximations in such an analysis are ones that
reflect the validity of the convolution framework itself,
as outlined in Sec.~\ref{ssec:wba}.
Namely, one assumes that within the WBA the form of the off-shell
nucleon function $\delta f^N$ remains the same for both $A=2$ and
$A=3$ nuclei, with the $A$ dependence of the off-shell structure
functions in Eqs.~(\ref{eq:conv_off}) entering only through the
off-shell smearing function $\widetilde{f}_{ij}^{\, N/A}$.
The model dependence of the smearing functions can be assessed
by considering different wave functions or spectral functions
for the deuteron and $A=3$ nuclei, as we have explored for the
KPSV and SS $^3$He spectral functions in Sec.~\ref{ssec:A3smear}.
Since the choice of wave function model is a discrete rather
than a continuous parameter, it is difficult to systematically
incorporate the uncertainty from this into the final error analysis.
The usual procedure is to examine the dependence of the results
on the choice of wave function and estimate the uncertainty from
the resulting variation.

\section{Conclusion}
\label{sec:conc}

With the completion of data taking in 2018 by the suite of $^3$He/$^3$H
experiments at Jefferson Lab, including MARATHON in DIS kinematics
\cite{MARATHON} and E12-11-112 in the QE region and beyond, there is
great anticipation to see the impact that the new data will have on
our knowledge of the structure of the free neutron, and in particular
on the $d/u$ PDF ratio at large $x$, which has eluded definitive
confirmation for over 3 decades.
Working within the weak binding approximation formalism, we have
revisited the calculation of deep-inelastic $A=3$ structure functions
using the latest theoretical developments, in terms of finite-$Q^2$
convolution formulas and nuclear effects computed from $A=3$ spectral
functions and off-shell nucleon structure functions.

To test the veracity of the WBA smearing functions and the range of
applicability of the impulse approximation, we have examined the
world's available data on inclusive $^3$He structure functions in
the vicinity of $x \approx 1$, which is expected to be dominated
by QE scattering.
Comparison with existing data from SLAC and Jefferson Lab suggests
that a good description can be obtained using the $Q^2$ dependent
smearing functions as for DIS, for $Q^2 \gtrsim 1$~GeV$^2$ out to
$x \approx 1.3$.
For smaller $Q^2$ rescattering and MEC effects are expected to be
more important, while for $x \gg 1$ the effects of the off-shell 
corrections and other multi-nucleon correlations will play a
greater role.
This analysis provides confidence in the application of the nuclear
model to the description of the $A=3$\ DIS and QE data, and suggests
that the extraction of neutron information from the MARATHON and
E12-11-112 Jefferson Lab data should not be impeded by the lack of
knowledge of the short-range structure of the $A=3$ wave functions.

For the bound nucleon structure functions, the WBA formalism allows
the inclusion of possible off-shell dependence in the calculation of
the nuclear structure functions, in both DIS and QE kinematics.
For the QE data comparisons, the off-shell corrections generally
improve the agreement between data and the WBA theory, especially
at low $Q^2$ values, irrespective of the prescription adopted for
the elastic off-shell nucleon structure function.
For the DIS region, the shape of the off-shell corrections for the
isoscalar channel, $\delta f^0$, is taken from the earlier CJ15
global QCD analysis of proton and deuterium data~\cite{CJ15} or
from the Kulagin-Petti fit to various nuclear structure function
ratios~\cite{KP06}, which assumed isospin symmetry for the
off-shell functions.

To explore possible constraints on the isospin dependence of the
off-shell functions, we performed a Monte Carlo analysis of the
recent data on the $^3$He/$d$ cross section ratios from the
Jefferson Lab E03-103 experiment~\cite{Seely09} for
$0.33 \lesssim x \lesssim 0.58$.
Within the statistical and systematic uncertainties of the data,
one can obtain almost equally good descriptions with no off-shell
corrections and with nonzero off-shell corrections with large
cancellations between the proton and neutron contributions
at large $x$.
The analysis disfavors, however, fits with nonzero off-shell
corrections which assume $\delta f^p = \delta f^n$.
Unfortunately, the results are quite sensitive to the shape of
the input isoscalar off-shell correction, and a robust analysis
must therefore involve a simultaneous fit to all proton, deuteron
and $A=3$ data.
Also, the lack of scatter of the E03-103 data points in
Fig.~\ref{fig:Seely} suggests that the data do not follow a
Gaussian probability distribution, so that the uncertainties
on the data points are dominated by systematic errors.
This highlights the important need for the new high-precision
data expected from the MARATHON experiment.

We have also outlined a new analysis procedure for extracting
the neutron structure function $F_2^n$ using Bayesian methods,
that would be capable of simultaneously extracting the free
neutron to proton structure function ratio and the proton
and neutron off-shell functions, $\delta f^p$ and $\delta f^n$,
within the nuclear WBA framework.
This would remove potential uncertainties in the extracted $F_2^n$
propagating from any assumptions made about the super-ratio,
${\cal R}$, of the $^3$He to $^3$H EMC ratios.
Instead, the method would utilize MARATHON data on the $^3$He/$d$
and $^3$H/$d$ ratios, as well as $d$/$p$ measurements to be used
for benchmarking against the global inclusive proton and deuteron
DIS data sets.
The MARATHON ratio data can also be supplemented with measurements of
the absolute values of the $F_2^p$ and $F_2^d$ structure functions
at similar kinematics in the E12-10-002 experiment~\cite{E12-10-002}
in Jefferson Lab's Hall~C, for $x \gtrsim 0.2$ and
$Q^2 \approx 5-16$~GeV$^2$.
The new data are eagerly awaited, and promise for the first time
to reveal the detailed structure of the free neutron at large $x$,
as well as the isospin dependence of the nuclear effects, and solve
the long-standing problem of the size of the nuclear EMC effect in
the deuteron.

\begin{acknowledgments}

We thank D.~Gaskell, C.~Keppel, S.~Li, H.~Liu, G.~Petratos and G.~Schnel
for helpful discussions and communications.
This work was supported by the U.S. Department of Energy contract
No.~DE-AC05-06OR23177, under which Jefferson Science Associates, LLC
operates Jefferson Lab.
A.T. was supported by NSF Grant No.~1359026 for an REU internship at 
ODU/Jefferson Lab, University Fellowship from the Ohio State Graduate
School, and the William A. Fowler Graduate Fellowship from the Ohio
State Department of Physics.
N.S. was supported by DOE Contract No.~DE-FG-04ER41309.

\end{acknowledgments}


\end{document}